\documentclass[aps,twocolumn,amsmath,amssymb,preprintnumbers,superscriptaddress,floatfix]{revtex4}
\usepackage[colorlinks,allcolors=blue]{hyperref}
\usepackage[capitalize]{cleveref} 
\usepackage{cleveref}

\usepackage{graphicx}

\renewcommand{\v}[1]{\boldsymbol{#1}}                           

\begin{document}
\title{Effective rheology of immiscible two-phase flow in porous media consisting of random mixtures of grains having two types of wetting properties}

\author{Hursanay Fyhn}\email{hursanay.fyhn@ntnu.no}
\affiliation{PoreLab, Department of Physics, Norwegian University of
  Science and Technology, NO--7491 Trondheim, Norway.}

\author{Santanu Sinha}\email{santanu.sinha@ntnu.no}
\affiliation{PoreLab, Department of Physics, University of Oslo,
  N--0316 Oslo, Norway}

\author{Alex Hansen}\email{alex.hansen@ntnu.no}
\affiliation{PoreLab, Department of Physics, Norwegian University of
  Science and Technology, NO--7491 Trondheim, Norway.}

\date{\today}
\begin{abstract}

We consider the effective rheology of immiscible two-phase flow in porous media consisting of random mixtures of two types of grains having different wetting properties using a dynamic pore network model under steady-state flow conditions. Two immiscible fluids, denoted by ``A'' and ``B'' flow through the pores between these two types of grains denoted by ``$+$'' and ``$-$''. Fluid ``A'' is fully wetting and ``B'' is fully non-wetting with respect to ``$+$'' grains whereas it is the opposite with ``$-$'' grains. The direction of the capillary forces in the links between two ``$+$'' grains is therefore opposite compared to the direction in the links between two ``$-$'' grains, whereas the capillary forces in the links between two opposite types of grains average to zero. For a window of grain occupation probability values, a percolating regime appears where there is a high probability of having connected paths with zero capillary forces. Due to these paths, no minimum threshold pressure is required to start a flow in this regime. While varying the pressure drop across the porous medium from low to high in this regime, the relation between the volumetric flow rate in the steady state and the pressure drop goes from being linear to a power law with exponent $2.56$ to linear again with increasing pressure drop. Outside the percolation regime, there is a threshold pressure necessary to start the flow. No linear regime is observed for low pressure drops. When the pressure drop is high enough for there to be flow, we find that the flow rate depends on the excess pressure drop to a power law with exponents around $2.2$--$2.3$. At even higher excess pressure drops, the relation becomes linear. We see no change in exponent for the intermediate regime at the percolation critical points where the zero-capillary force paths disappear. We measure the mobility at the percolation threshold at low pressure drops so that the flow rate versus pressure drop is linear. Assuming a power law, the mobility is proportional to the difference between the occupation probability and the critical occupation probability to a power of around $5.7$. 
\end{abstract}
\maketitle

\section{Introduction}%
\label{sec:introduction}

It was in 1827 that Ohm published his law stating that electrical current is proportional to the voltage drop across a conductor \cite{o27}, meeting fierce resistance from the physics community in the beginning. Darcy arrived in 1856 at a similar law for single-phase flow in porous media, i.e., the volumetric flow rate is proportional to the pressure drop across the porous medium \cite{d56}.  Both of these fundamental laws are examples of there being a linear relationship between current and driving force. In the case of the Darcy law, the derivation based on pore scale physics has been a challenge, see e.g., Whitaker's derivation based on momentum transfer \cite{w86}. 

The Darcy law for single-phase flow through a porous sample is 
\begin{equation}
\label{eqah1}
Q=-\frac{AK}{\mu L}\ \Delta P\;,
\end{equation}
where $Q$ is the volumetric flow rate along the axis of the cylindrical sample, $\Delta P$ the pressure drop along it in the flow direction, $A$ the area of the sample orthogonal to the flow direction, $K$ the permeability of the sample, $\mu$ the viscosity of the liquid and $L$ is the system length.  

In 1936 the Darcy law (\ref{eqah1}) was generalized to the simultaneous flow of two immiscible liquids by Wyckoff and Botset by essentially splitting it into two \cite{wb36},
\begin{eqnarray}
Q_w=-\frac{AKk_{rw}}{\mu_w L}\ \Delta P\;,\label{eqah2}\\
Q_n=-\frac{AKk_{rn}}{\mu_n L}\ \Delta P\;,\label{eqah3}\\
\nonumber
\end{eqnarray}
where the subscripts $w$ and $n$ refer to the wetting properties of the two fluids with respect to the matrix; $w$ refers to the more wetting fluid and $n$ to the less wetting fluid.  The idea behind this split is simple.  The wetting fluid will see a pore space reduced by the presence of the other fluid, leading to a reduction in effective permeability for the wetting fluid.  The reduction parameter is the wetting relative permeability $k_{rw}$.  Completely analogously the non-wetting fluid sees an effective reduction of the permeability by a factor $k_{rn}$, the non-wetting relative permeability.  The split was given physical contents when Wyckoff and Botset assumed that the two relative permeabilities were functions of the wetting saturation $S_w$ alone; the wetting saturation being the pore volume occupied by the wetting fluid divided by the total pore volume and assuming the fluids are incompressible. Barenblatt et al.\  
\cite{bps02} have later shown that this assumption is valid if there exists a local phase equilibrium between the fluids, a condition that is fulfilled only for slow flows. A further assumption built into equations (\ref{eqah2}) and (\ref{eqah3}) is that there are no macroscopic saturation gradients present. 

The total volumetric flow rate is given by the sum of the volumetric flow rates of each fluid, 
\begin{equation}
    \label{eqah4}
    Q=Q_w+Q_n\;,
\end{equation}
and as a consequence, the generalized Darcy equations (\ref{eqah2}) and (\ref{eqah3}) predict 
\begin{equation}
    \label{eqah6}
    Q=- \frac{AK}{L} \left[ \frac{k_{rw}}{\mu_w}+\frac{k_{rn}}{\mu_n}\right]\ \Delta P\;, 
\end{equation}
that is, a total volumetric flow rate being proportional to the pressure drop.

Equations (\ref{eqah2}) and (\ref{eqah3}) assume that there are no macroscopic saturation gradients.  If this is not the case, the pressure is split into one associated with the non-wetting fluid, $P_n$, and one associated with the wetting fluid, $P_w$. Their difference is equal to the capillary pressure function, $P_n-P_w=P_c(S_w)$, which is also assumed only to depend on the saturation $S_w$. Equations (\ref{eqah2}) and (\ref{eqah3}) will then contain terms of the type $\nabla P_c=(dP_c/dS_w)\nabla S_w$, thus setting up the pressure gradient and the saturation gradient as driving forces. When these equations are combined with mass conservation, the result is a closed set of equations that determine how the saturation develops within the porous medium.

When the saturation changes inhomogeneously in the porous medium with time,   
one implicitly assumes that fluid interfaces move within the porous medium.
It was then a surprise when Tallakstad et al.\ \cite{tallakstad2009steady,tallakstad2009steadyE} reported a flow rate $Q$ depending on $\Delta P$
as
\begin{equation}
    \label{eqah7}
    Q\propto|\Delta P|^\beta\;,
\end{equation}
with $\beta \approx 1.85$ for a two-dimensional glass-bead-filled Hele-Shaw cell filled with a water-glycerol mixture and air in the flow regime where the generalized Darcy equations (\ref{eqah2}) and (\ref{eqah3}) are supposed to be valid. This study was followed up by an NMR study of the three-dimensional glass bead packings by Rassi et al.\ \cite{rcs11} finding an exponent $\beta$ 
varying between 2.2 and 3.3.   Aursj{\o} et al. \cite{aetfhm14} using the same model porous medium
as Tallakstad et al.\ \cite{tallakstad2009steady,tallakstad2009steadyE}, but with two incompressible fluids, found $\beta \approx 1.5$ or $1.35$ depending on the fractional flow rates.    Similar results in the sense that $\beta$ is considerably larger than one, have since been observed by a 
number of groups, see \cite{sh17,glbb20,zbglb21,zbb22}. There has also been a considerable effort to
understand these results theoretically and reproduce them numerically  
\cite{tallakstad2009steady,tallakstad2009steadyE,gh11,sh12,shbk13,xw14,ydkst19,rhs19,rsh19,fyhn2021rheology,ssoa22,ffh22,lhrt22,cfhws23}. 

It should be pointed out that the power law behavior seen in Eq.\ (\ref{eqah7}) is different from the one described by Wilkinson in 1986 \cite{wilkinson86}. In that work, Wilkinson used the invasion percolation model to work out the dependence of the relative permeabilities on the capillary pressure, which could be linked to the saturation.  He would find that the non-wetting relative permeability $k_{rn}$ would depend on the difference between the capillary pressure $P_c$ and a critical capillary pressure $P^c_c$ related to the percolation critical point as
\begin{equation}
\label{eqah7-1}
k_{rn}\sim (P_c-P^c_c)^t\;,
\end{equation}
where $t$ is the percolation conduction exponent \cite{sa18}.  This is, however, a very different problem from the one giving rise to Eq.\ (\ref{eqah7}). The power law in (\ref{eqah7-1}) is a direct reflection of the geometry of the clusters of the non-wetting fluid in the system {\it after\/} the invasion process. Hence, it is a {\it static\/} problem.  The power law in (\ref{eqah7}) is, as we shall see, the result of a {\it dynamic\/} process caused by the motion of the fluid interfaces.   

The power law behavior in equation (\ref{eqah7}) is due to a competition between the capillary and the viscous forces.  It is straightforward to understand why the flow rate should increase faster than linear when these forces are in competition. When the pressure difference across the porous medium is increased, more interfaces are beginning to move leading to a higher effective permeability \cite{rh87}. Why it should be a power law is less obvious.  The best argument for why was perhaps already given by Tallakstad et al.\ \cite{tallakstad2009steady,tallakstad2009steadyE} through comparing pressure drop across fluid clusters with the capillary pressures holding them in place. Capillary fiber bundle models \cite{s53,s20} are porous media in the form of bundles of capillary fibers and they are typically simple enough to be mathematically solvable 
\cite{shbk13,rhs19,rsh19,fyhn2021rheology,lhrt22,cfhws23}.   When the fibers have undulating radii along the long axis, they show non-linear volumetric flow rate vs.\ pressure drop, but not quite of the form (\ref{eqah7}), but rather 
\begin{equation}
    \label{eqah8}
    Q=\ \left\{  
    \begin{array}{ll}
         0 &  \mbox{if $|\Delta P| \le P_t$}\;,\\ 
        M(|\Delta P|-P_t)^\beta &  \mbox{if $P_t < |\Delta P| <P_{\max}$}\;,\\
         M_D(|\Delta P|-P_t) &  \mbox{if $P_ {\max} \ll |\Delta P|$}\;,
    \end{array}                
    \right.  
\end{equation}
where $P_t$ is a threshold pressure necessary for flow to occur, $P_{\max}$ is the maximum threshold pressure found in any capillary fiber, and  $M$ and $M_D$ are mobilities.  
A non-zero threshold pressure is in general necessary in porous media when neither of the two immiscible fluids percolates when dealing with porous media and not just the capillary fiber bundle model \cite{sh12,sh17,fyhn2021rheology}.  
The existence of a non-zero threshold pressure makes the measurement of $\beta$ much harder than when it is zero as this implies determining two parameters simultaneously, $(P_t,\beta)$, rather than just one, $\beta$. 

A central unanswered question is whether the exponent $\beta$ is universal in the sense that there are classes of systems that all have the same value, i.e., can one define universality classes?  Intuitively this is a very appealing idea as one has a diverging length scale as in equilibrium critical phenomena as $|\Delta P|\to P_t$ from above
\cite{tallakstad2009steady,tallakstad2009steadyE}.  The experimental measurements of $\beta$ have so far not given any indication of the existence of universality classes and neither have the computational efforts due to the difficulties in dealing with two unknown parameters, $P_t$ and $\beta$. Roy et al.\ \cite{rhs19} found using a capillary fiber bundle model that $\beta=2$ if the fiber-to-fiber probability distribution of thresholds includes $P_t=0$ with a finite probability, otherwise $\beta=3/2$. The fibers here had smoothly undulating radii along the flow direction.  Lanza et al.\ \cite{lhrt22} who studied non-Newtonian a mixture of immiscible Newtonian and non-Newtonian fluids in a capillary fiber bundle model found a different value of $\beta$ when the radius distribution is jagged from when it is smooth. 

Equation (\ref{eqah8}) which was derived for the capillary fiber bundle model, predicts there being a pressure drop $|\Delta P|=P_t$ below which the flow rate $Q$ is zero. This threshold may be zero. Within the capillary fiber bundle model, this means that some capillary tubes belonging to the bundle have interfaces that move as soon as there is a pressure difference across them.  There is, however, one important mechanism missing in the capillary fiber bundle models: In the porous medium, the immiscible fluids may be percolating. That is, there are pathways through the porous medium along which there are no interfaces. In this case, there will be a {\it linear\/} regime when the pressure drop is low enough so that the interfaces surrounding the percolating paths do not move.  When the pressure drop is increased sufficiently for them to do so, the non-linear power law regime sets in.

Recently Fyhn et al.\ \cite{fyhn2021rheology} studied the exponent $\beta$ and the threshold pressure $P_t$ in a capillary fiber bundle model and a dynamic pore network model under mixed wet conditions.  In the dynamic pore network model, each link was given a wetting angle --- in the sense that if there is an interface in the link, this is the angle it will make with the walls of the tube --- drawn from a given probability distribution.  In the capillary fiber bundle model, each undulating tube is given a wetting angle from a given probability distribution. In both models, a constitutive law of the form (\ref{eqah8}) was found. The capillary fiber bundle model could be solved analytically, yielding
\begin{equation}
\label{eqah9}
   \beta=\ \left\{  
    \begin{array}{ll}
         1   &  \mbox{if $|\Delta P| - P_t\gg P_{\max}$}\;,\\ 
         2   &  \mbox{if $P_t \ll |\Delta P| - P_t\ll P_{\max}$}\;,\\ 
         3/2 &  \mbox{if $0 < |\Delta P| - P_t\ll P_t$}\;. 
    \end{array}                
    \right.  
\end{equation}
The network model studies showed a less clear picture, with $\beta$ varying between 1 and 1.8 depending on the saturation and the wetting angle distribution.  It was not possible to resolve whether there were regions of fixed $\beta$ or whether it varied continuously with the parameters of the model.  This was due to the non-zero threshold pressure $P_t$ which needed to be determined together with $\beta$.

We study here a model for immiscible two-phase flow in a porous medium made from two types of grains that have different wetting properties with respect to the fluids.  The model treats the interfacial tension between the two fluids similarly to a model introduced by Irannezhad et al.~\cite{ipbrz22,ipbrz23}. 
We imagine a packing of two types of grains, say type ``$+$'' and type ``$-$''.  Two immiscible fluids, denoted by ``A'' and ``B'' flow through the pores between the grains denoted by ``$+$'' or ``$-$''.  Fluid ``A'' is fully wetting and ``B'' is fully non-wetting with respect to ``$+$'' grains whereas it is the opposite with respect to ``$-$'' grains. The direction of the capillary forces in the links between two ``$+$'' grains is therefore opposite compared to the direction in the links between two ``$-$'' grains, whereas {\it the capillary forces in the links between two opposite types of grains are zero.\/}  

\begin{figure}[ht]
  \centering
  \includegraphics[width=0.9\linewidth]{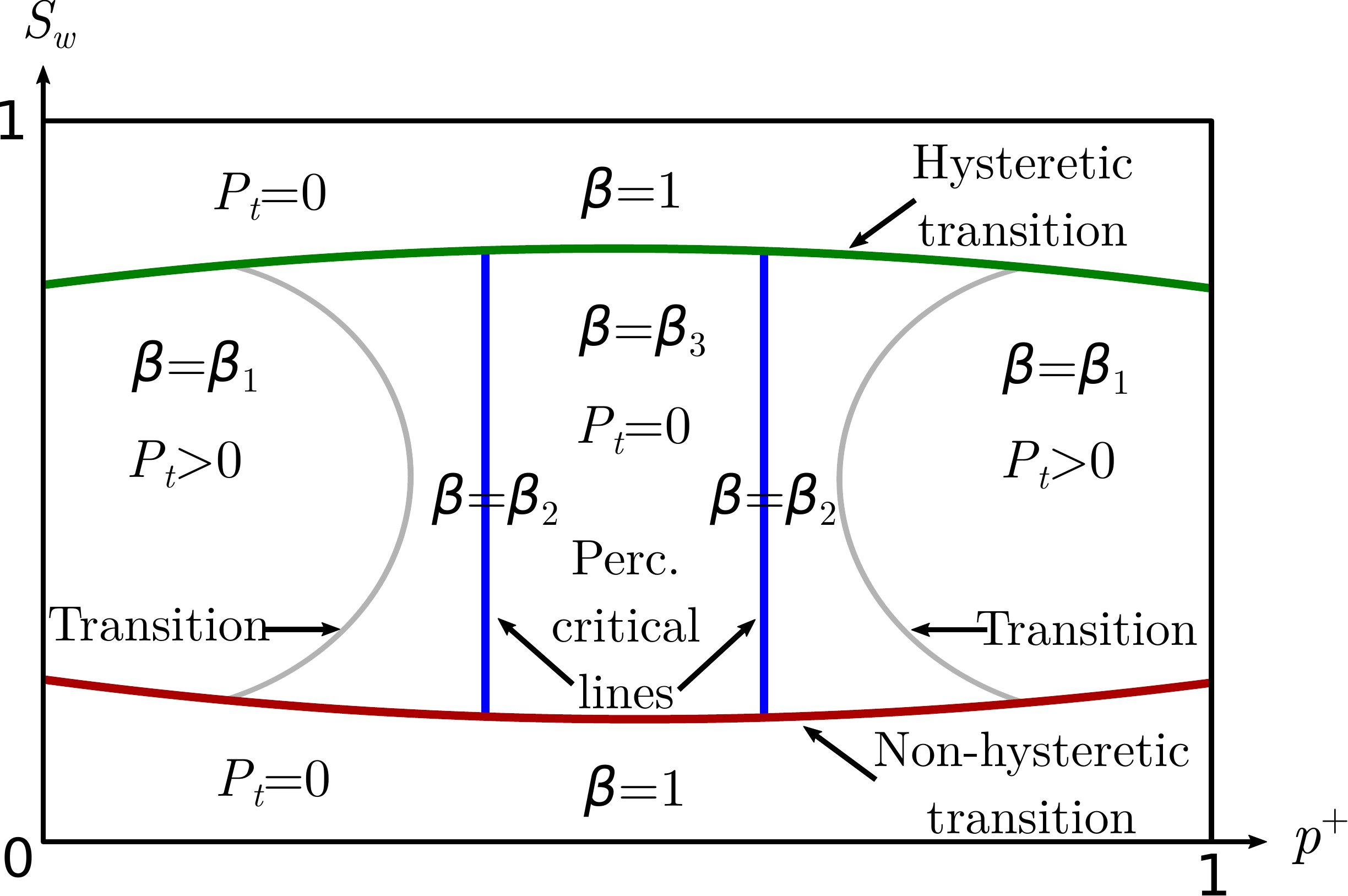}
  \caption{Phase diagram showing the exponent $\beta$ and the threshold pressure $P_t$
  plotted against the occupation probability $p^+$ and the saturation $S_w$. The diagram is symmetric about the $p^+=1/2$ line.  The two vertical blue lines are critical lines associated with the two percolation transitions.  The lower red line is separating two-phase flow from single-phase flow. There is no hysteresis associated with this line.  The upper green line also distinguishes between two-phase and single-phase flow.  However, in this case, there is hysteresis. The two gray lines represent the transition lines from threshold pressure $P_t=0$ to a non-zero value.  The nature of these lines is unknown.  
  }%
  \label{fig1}
\end{figure}

The probability that a grain is of ``$+$'' type, is $p^+$.  A second parameter is the wetting saturation $S_w$.  There is a rich phase diagram when plotting the threshold pressure $P_t$ as a function of the two control variables $p^+$ and $S_w$ which is illustrated in Fig.\ \ref{fig1}. 
Note in particular in this phase diagram that there is a region in the middle where the threshold pressure $P_t=0$.  This region is limited by two $p^+={\rm constant}$ critical lines.  Each line signifies a percolation transition \cite{sa18}. The two curved grey lines signify a possible shift of the two blue transition lines due to the dynamics of the model.   
There are also two other lines, one green line marked ``hysteretic transition'' and one red line marked ``non-hysteretic'' transition.  Crossing such a line, one of the two fluids stops moving and we are essentially dealing with a single-phase flow problem. When the wetting fluid stops moving, there is no hysteresis. On the other hand, when the non-wetting fluid stops, there is hysteresis in the sense that the wetting saturation has to be lowered more to get the non-wetting fluids moving again \cite{kh06}.  

In the region where $P_t=0$, while still having two-phase flow, we observe an exponent $\beta=\beta_3=2.56\pm 0.05$ for saturation $S_w=0.5$.
This value is also seen when setting $p^+=p_c$ or $p^+=1-p_c$, where $p_c\approx 0.5927$ is the site percolation threshold for square lattices,  i.e., $\beta=\beta_2=\beta_3$. For $p^+$ much lower than $1-p_c$ or $p^+$ much higher than $p_c$, still for saturation $S_w=0.5$, we see $\beta=\beta_1=2.25\pm0.1$. The large uncertainty in $\beta$ seen here stems from $P_t>0$.    

It is surprising that $\beta_2=\beta_3$ within the precision we are able to obtain.  The exponent $\beta_2$ is obtained at the percolation threshold where the paths where fluid surfaces meet no resistance are fractal with fractal dimension 4/3 as they are the external perimeters of percolation clusters \cite{ga86}. The reason for not seeing critical behavior reflected in $\beta$ comes from there also being other links that have no interfacial tension in them as they contain no interfaces, thus driving the system away from criticality.  In order to investigate whether there are any traces at all in the transport properties of the percolation critical point, we have studied the mobility $M$ at low pressure drops as $p^+$ approaches a critical value, where we expect it to vanish with an exponent $t'$ of the same type as the conductivity exponent $t$ percolation in ordinary percolation in the vicinity of the critical $p^+$ \cite{sa18,r07}. We find that $t' \approx 5.7$, indicating that the system is {\it not\/} critical and $M$  falls off faster than algebraic. We therefore expect there to be two extra transition lines (marked in grey in Fig.\ \ref{fig1}) that distinguish between $P_t=0$ and $P_t>0$.  The nature of these lines is unknown. 

We use a dynamic pore network model \cite{sinha2019dynamic,amhb98,kah02,toh12,gvkh18} for this study. It has been used earlier in the context of modeling mixed wetting porous media, see \citep{sinhaGrova2011local,flovikSinha2015dynamic,fyhn2021rheology}. We describe the model in Section \ref{DPN} including our use of the wetting model similar to the one introduced by Irannezhad et al.\ \cite{ipbrz22,ipbrz23}.  Section \ref{easy} explains how we identify the paths through the network that have no capillary forces associated with them and relate them to a site percolation problem.  Section \ref{mobility} presents the analysis of the low pressure drop mobility at the percolation critical points. Section \ref{non-darcy} constitutes our investigation of volumetric flow rate $Q$ vs.\ pressure drop $\Delta P$.  We fix the saturation $S_w=0.5$ and scan through this line in the phase diagram in Fig.\ \ref{fig1} for different values of $p^+$. We also tested whether there would be hysteresis with respect to increasing or decreasing the pressure drop, finding none.  Section \ref{sec:conclusion} contains a summary and our conclusions.

\section{Dynamic Pore Network Model}%
\label{DPN}

\begin{figure}[ht]
  \centering
  \includegraphics[width=1.0\linewidth]{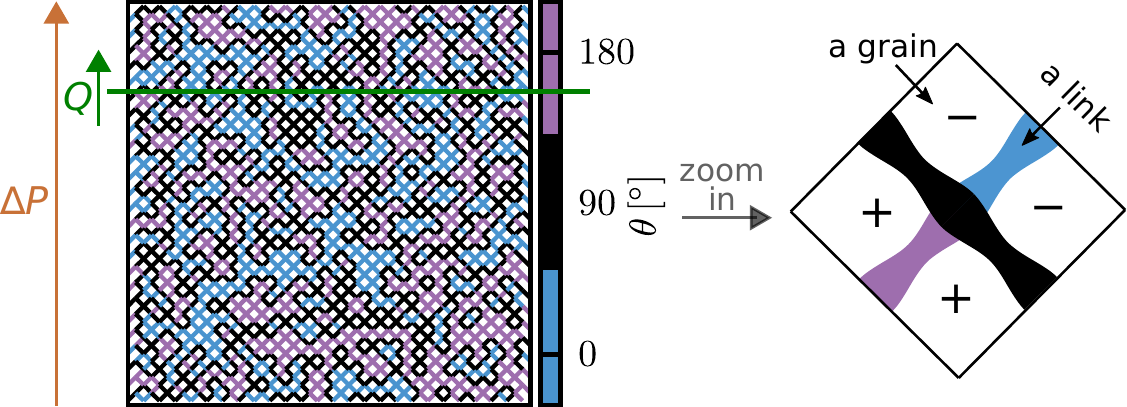}
  \caption{The dynamic pore network model implemented on a square lattice consists of links oriented $45^{\circ}$ from the overall flow direction.
  The flow is driven by a global applied pressure $\Delta P$ and the total volumetric flow rate $Q$ is measured over a cross-section normal to the direction of the overall flow. The wetting angle $\theta$ of each link is based on its adjacent grains.
  The grains are assigned ``$+$'' with an occupation probability $p^+$ and the rest of the grains are assigned ``$-$''.  
  If both of the adjacent grains are assigned ``$+$'', $\theta = 180^\circ$ (marked pink).
  If both of the adjacent grains are assigned ``$-$'', $\theta = 0^\circ$ (marked blue).
  Lastly, if one of the adjacent grains is ``$+$'' and the other one is ``$-$'', then $\theta = 90^\circ$ (marked black) and hence there are no capillary forces associated with interfaces in the link.
  }%
  \label{fig:DPN_sketch}
\end{figure}

A sketch of the dynamic pore network (DPN) model used in this work is given in \cref{fig:DPN_sketch}, showing a square two-dimensional network with links with the same length tilted $45^\circ$ angle from the flow direction. The $\Delta P$ across the network drives the flow leading to a $Q$ which is measured over a cross-section of the system normal to the direction of the overall flow. 
The zoomed-in sketch to the right in \cref{fig:DPN_sketch} illustrates the rules for using the wetting properties of the grains to assign wetting angles $\theta$ to the links, where $\theta$ is consistently defined through one of the fluids.
In contrast to earlier models \cite{sinhaGrova2011local,flovikSinha2015dynamic,fyhn2021rheology} that assign the wetting angles to the pores or links directly, the physical basis for this model is a mixture of grains and the wettability of the pore space in-between depends on the wettability of the surrounding grains, similar to the system introduced by Irannezhad et al.\ \cite{ipbrz22,ipbrz23}.
We assume two types of grains, they being either fully non-wetting with $\theta = 180^\circ$ or fully wetting with $\theta = 0^\circ$.
Having fully non-wetting or fully-wetting grains maximizes the difference between the two types of grains in terms of their wettability and hence maximizes any impact on the rheology that comes as a result of this difference.
The grains are denoted fully non-wetting and assigned a notation ``$+$'' with an occupation probability $p^+$ and the rest of the grains are then fully wetting with a notation ``$-$''.  For each link, $\theta$ is determined based on the link's adjacent grains.
Each grain in the network is connected to four links which means each link has two adjacent grains, as shown in \cref{fig:DPN_sketch}.
If both of the adjacent grains are assigned ``$+$'', the link will have $\theta = 180^\circ$.
If both of the adjacent grains are assigned ``$-$'', then $\theta = 0^\circ$.
Lastly, if one of the adjacent grains is ``$+$'' and the other one is ``$-$'', the link in the middle should be easy to pass through for both fluids and the wettability should be neutral with $\theta = 90^\circ$.

The networks have periodic boundary conditions in both directions.
Two fluids that flow through the network are immiscible and their movement is traced through the position of their interfaces at each instant in time. 
Whenever the fluids flowing in a link reach the crossing point with the three other links, namely a node, the fluids get distributed into the neighboring links in the same time step instead of being retained in the node itself \cite{sinha2019dynamic}.

\begin{figure}[ht]
  \centering
  \includegraphics[width=0.7\linewidth]{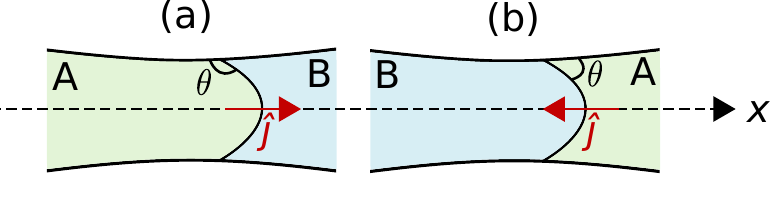}
  \caption{
  The wetting angle $\theta$ is consistently measured through fluid~A in both examples (a) and (b) regardless of the wettability situation. 
  The unit vector $\hat J$ lies along the center axis $\v x$, and points in the direction out of the fluid within which $\theta$ is measured, which in this case is from fluid~A to fluid~B.
  }%
  \label{fig:theta_sketch}
\end{figure}

The volumetric flow in each link with length $l$, pointing along the link's center-axis $\v x$, is given by
\begin{equation}
  \label{eq:q}
  q =
  - \frac{\pi \bar r ^4}{8\mu l} \left(\Delta p - \hat x \cdot \left(\sum_k p_t(x_k) \hat J_k \right)\right),
\end{equation}
where it has been assumed that the radius does not deviate too much from its average value $\bar r$ \cite{sinha2019dynamic}.
Here, $\mu = s_A\mu_{A} + s_B\mu_B$ is the saturation weighted viscosity of the fluids where $s_A$ and $s_B$ are saturations of the two fluids $A$ and $B$ respectively with viscosities $\mu_A$ and $\mu_B$ in the links (in contrast to $S_w$ which is the average saturation over the whole network).
\Cref{fig:theta_sketch} can be used to further explain the variables in \cref{eq:q}.
The unit vector $\hat J$ lies along the $\v x$ axis, and points in the direction \textit{out of} the fluid within which $\theta$ is defined.
In \cref{fig:theta_sketch}, $\theta$ is consistently measured through fluid~A in both examples (a) and (b) regardless of if fluid~A is more or less wetting with respect to the solid. 
Hence, $\hat J$ also consistently points across the interface starting from fluid~A towards fluid~B.
The sum in \cref{eq:q} is taken over the interfaces numbered $k$ with varying $\hat J_k$ with positions $x_k \in [0,l]$ along $\v x$.
The dot product of this sum with the unit vector $\hat x$ in the positive $\v x$ direction is taken afterward to obtain the total capillary pressure.
The capillary pressure across one interface at position $x$ which has an angle $\theta$ with the solid through fluid~A is modeled by using the Young-Laplace equation~\cite{blunt2017multiphase},
\begin{equation}
  \label{eq:pcapillary}
  p_t(x) = \frac{2\sigma \cos{\theta}}{r(x)},
\end{equation}
where $\sigma$ is the surface tension and
\begin{equation}
  \label{eq:radius}
  r(x) = \frac{r_0}{1-a\cos\left( \frac{2\pi x}{l} \right)}  
\end{equation} 
is the radius where $a$ is the amplitude of the periodic variation and $r_0/a$ is randomly chosen from the interval $[0.1 l, 0.4 l]$.
This way, $p_t$ varies with both the position along a link and from link to link.

For all simulations in the following, the two immiscible fluids have been given surface tension $3.0\cdot 10^{-5}$~N$/$mm and viscosity $0.1$~Pa$\cdot$s for both.
The overall network saturation is kept constant at $0.5$, meaning there is equal amounts of the two fluids.  The links in the network have length $l=1$~mm.
In all the figures, the logarithms are in base 10.

\section{Easy links and connected paths}
\label{easy}

\begin{figure}[ht]
  \centering
  \includegraphics[width=0.8\linewidth]{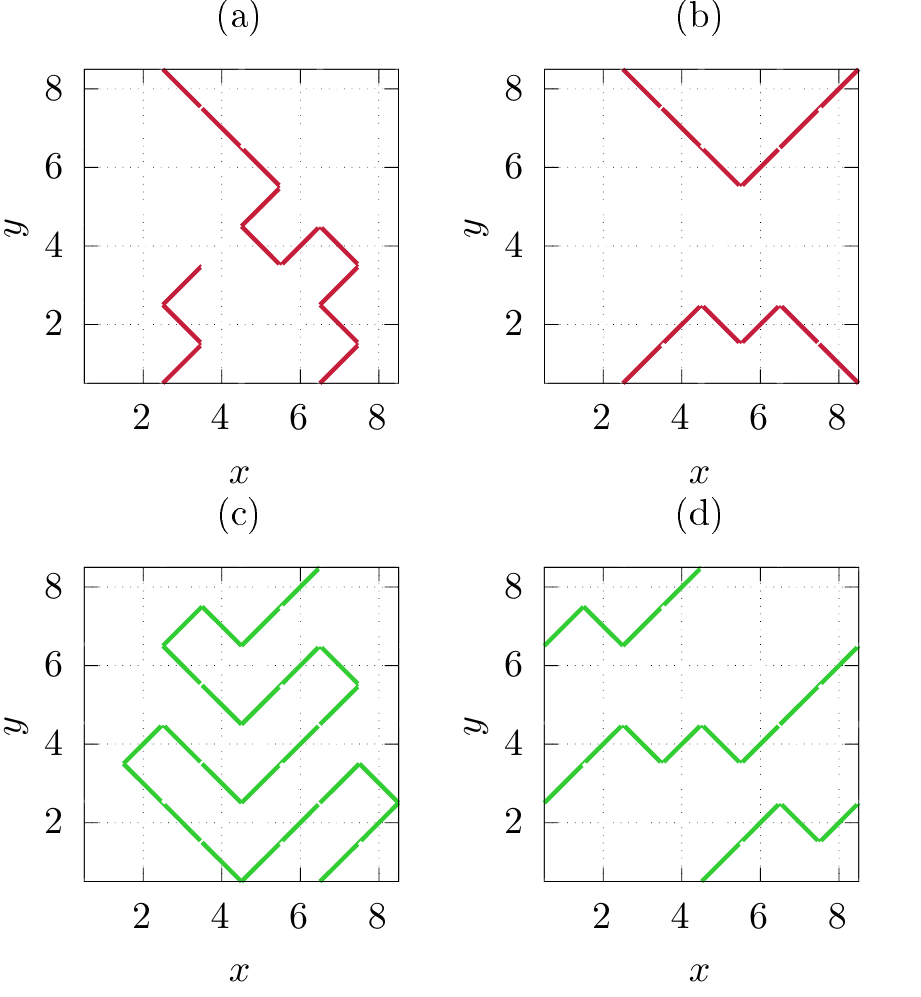}
  \caption{Due to periodic boundary conditions in both directions parallel and orthogonal to $Q$, it is not enough for a path to connect the bottom to the top of the network in the direction of $Q$ to qualify as a connected path, it also has to loop back to itself. We show here four examples of clusters of easy links in an $8\times 8$ lattice. Figures (a) and (b) do not qualify as connected paths as defined for the networks in this work, while (c) and (d) do. The cluster of easy links in  (a) connects the top and the bottom of the network but needs one additional link centered at $(x,y)=(4,4)$ to form a connected path. 
  The cluster of easy links in (b) forms a closed loop but does not cross the network fully in the flow direction which is along the $y$-axis. 
  In (c), the link centered at position $(x,y)=(1,7)$ meets the link centered at $(x,y)=(6,8)$ due to the periodic boundary condition in the $y$-direction and completes the loop, hence making the path a connected path.
  The effect of having periodic boundary condition in the $x$-direction is apparent in (d), where the link centered at $(x,y)=(8,2)$ connects to the link at $(x,y)=(1,3)$, and similarly, $(x,y)=(8,6)$ connects to $(x,y)=(1,7)$ and $(x,y)=(4,8)$ connects to $(x,y)=(5,1)$, thus completing the loop.}%
  \label{fig:sketchConPath}
\end{figure}

There are three types of links in the model: those that are of the ``$+$ $+$'' type, those that are of the ``$-$ $-$'' type, and the ``$+$ $-$''$=$``$-$ $+$'' type. We will in the following refer to the latter type as ``easy links'' since they offer no capillary resistance to interfaces that happen to be in them. Paths of connected easy links may percolate, i.e., stretch across the network forming loops as we are implementing bi-periodic boundary conditions.  We will refer to such percolating paths of easy links as ``connected paths'', see Fig.\ \ref{fig:sketchConPath}.

\begin{figure}[ht]
  \centering
  \includegraphics[width=1.0\linewidth]{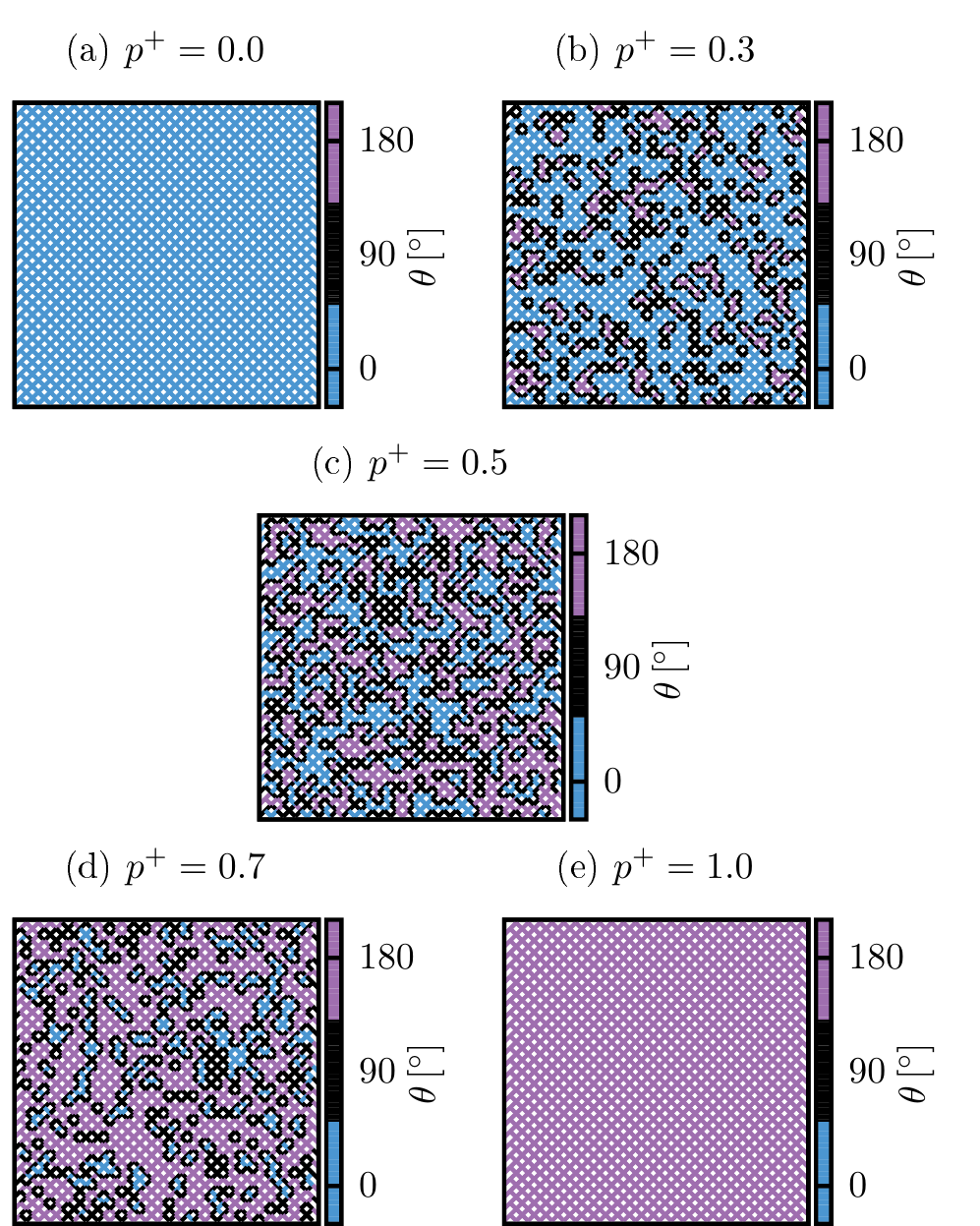}
  \caption{At $p^+ = 0$, shown in (a), all the grains are fully wetting grains that are noted as ``$-$'' in \cref{fig:DPN_sketch}. This means that the pore space between these grains, namely the links in DPN, all have $\theta = 0 ^\circ$. Oppositely at $p^+ = 1.0$, shown in (e), there are only links that have $\theta = 180 ^\circ$. In these two extreme cases, there is no easy link in the network with neutral-wettability $\theta = 90 ^\circ$.
  Moving away from these extremes, when $p^+ = 0.3$ in (b) or when $p^+ = 0.7$ in (d), links with $\theta = 90 ^\circ$ are present but not enough to create a connected path that crosses the entire system. At the middle point of $p^+ = 0.5$,(c), DPN has half of each type of grain creating the highest possible probability for having connected paths with only $\theta = 90 ^\circ$ links. In these examples, $p^+ = 0.5$ (c) is the only one that lies within the limit $1- p_c < p^+ < p_c$, and it is only here we find connected paths.
}%
  \label{fig:thetamap}
\end{figure}

The geometry of the easy links and connected paths may be mapped onto an ordinary site percolation problem \cite{sa18}. The links altogether form a square lattice. 
The nodes of the {\it dual\/} lattice, form another square lattice \cite{s77} and are assigned ``$+$'' or ``$-$''. 
These values are placed at random.  The distribution of neighboring ``$+$'' sites in this dual lattice form an ordinary site percolation problem. In an infinitely large lattice, there will be a percolating ``$+$'' cluster when $p^+\ge p_c$, where $p_c$ is the site percolation threshold $0.5927\dots$.
If we, on the other hand, focus on the ``$-$'' sites, there will be a cluster of such sites that percolate if $p^-=1-p^+\ge p_c$, or $p^+\le 1-p_c\approx 0.4073\dots$ \cite{xhz21}. 
Hence, if $0\le p^+\le 1-p_c$, the ``$-$'' clusters percolate, if $1-p_c\le p^+ \le p_c$, neither the ``$-$'' sites nor the ``$+$'' sites percolate, and if $p_c \le p^+$, the ``$+$'' sites percolate. 
We show in Fig.\ \ref{fig:thetamap} a map of the wetting angles associated with different values of $p^+$.  The easy links are shown in black.  

We note that if neither the ``$+$'' sites nor the ``$-$'' sites percolate ($1-p_c\le p^+ \le p_c$), there must be connected paths. We furthermore note that if either of the two site types percolates, there cannot be any connected paths.  At the two thresholds, $p^+=1-p_c$ and  $p^+=p_c$, the connected paths appear together with the appearance of a percolating cluster of either ``$-$'' or ``$+$'' type, as the perimeter of the incipient percolating cluster is a connected path.  
At the percolation thresholds, we know that the fractal dimension of the perimeter, and hence the corresponding connected path, is $4/3$ \cite{ga86}.
For values away from the critical points, the connected paths are not fractal.  Hence, the structure of the easy link clusters and the connected path is very different away from the critical points, while still being in the interval $1-p_c\le p^+ \le p_c$.

\begin{figure}[ht]
  \centering
  \includegraphics[width=1.0\linewidth]{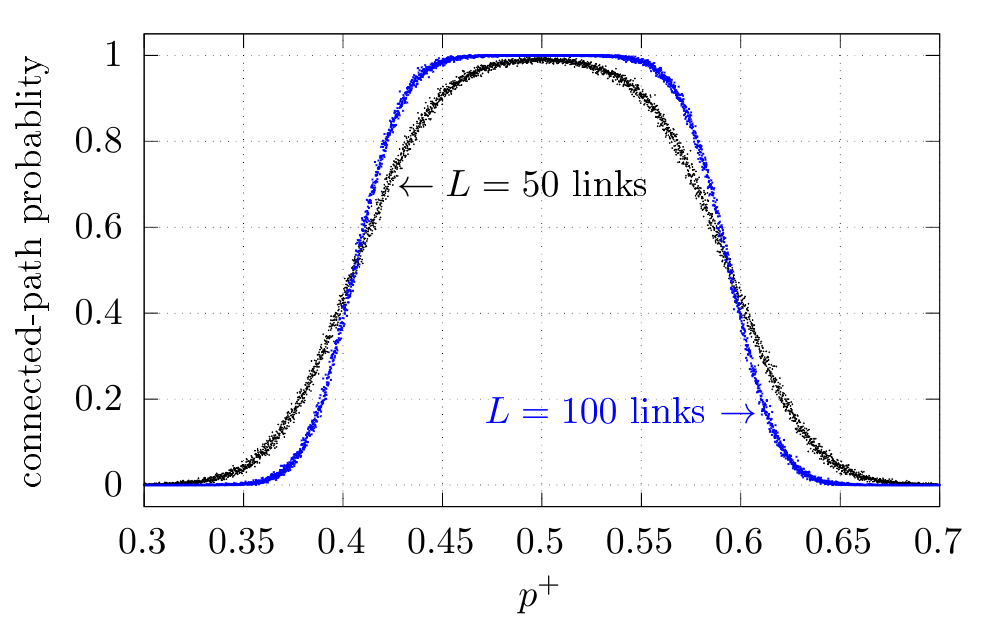}
  \caption{Probability for having connected paths in systems with non-wetting grain probability $p^+$ and size $L\times L$, where $L$ is either 50 or 100 links.}%
  \label{fig:connectedPath}
\end{figure}

The probability of finding a connected path as a function of $p^+$ is investigated by testing $1000$ randomly generated networks with size $L\times L$ for each $p^+ \in \{0.3000,0.3001,0.3002,\dots ,0.7000\}$. 
The results are shown in \cref{fig:connectedPath} for $L = 50$~links and $L = 100$~links. We see that the two curves cross very close to $1-p_c$ and $p_c$. 

\section{Mobility}
\label{mobility}

As we will show with the results presented in the next section, the constitutive law between the volumetric flow rate $Q$ and the pressure drop $|\Delta P|$ can be written as
\begin{equation}
    \label{eqah80}
    Q=\ \left\{  
    \begin{array}{ll}
        M|\Delta P| &  \mbox{if $|\Delta P| < P_l$}\;,\\ 
        M_m|\Delta P|^{\beta_3} &  \mbox{if $P_l< |\Delta P|<P_u$}\;,\\
        M_D|\Delta P| &  \mbox{if $P_u<|\Delta P|$}\;,\\
    \end{array}                
    \right.  
\end{equation}
in the region $1-p_c \le p^+ \le p_c$.
Here, $P_l$ and $P_u$ are two crossover pressures.  There are three regimes: (1) a linear regime at low pressure drop, (2) a non-linear regime for intermediate pressure drops and (3) a linear regime for high pressure drops.  
Each regime is characterized by a mobility, $M(p^+,S_w)$, $M_m(p^+,S_w)$, and $M_D(p^+,S_w)$, respectively.  

If we move to values of $p^+$ where $P_t>0$, regime (1) disappears.  Hence, we have that $M(p^+,S_w)$ tends to zero as $p^+$ reaches the boundary between the $P_t=0$ region and the $P_t>0$ region.  We hypothesize in the following that the boundaries of this region are given by the percolation thresholds $1-p_c$ and $p_c$.

\begin{figure}[ht]
  \centering
  \includegraphics[width=1.0\linewidth]{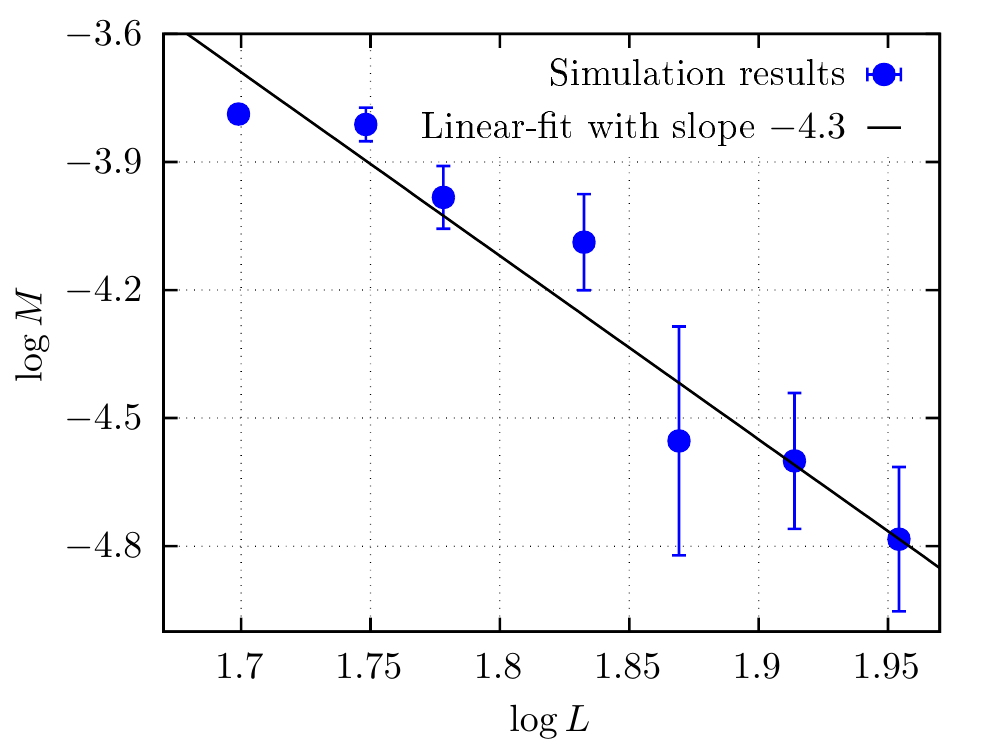}
    \caption{Mobility $M$ in networks with size $L\times L$. The slope of the linear fit is $- t'/\nu = -4.3\pm 1.0$. The saturation was set to $S_w=0.5$ in this calculation.}%
  \label{fig:logL_logM}
\end{figure}

Expecting that $M(p^+,S_w)$ shows similar behavior to the conductivity in percolation  
\cite{r07}, we make the assumption that the mobility vanishes as
\begin{equation}
    \label{eqah81}
    M \sim\ \left\{  
    \begin{array}{ll}
        (p^+-(1-p_c))^{t'} & \mbox{for $p^+\to (1-p_c)^+$}\;,\\ 
        (p_c-p^+)^{t'} &  \mbox{for $p^+ \to (p_c)^-$}\;,\\
    \end{array}                
    \right.  
\end{equation}
where $t'$ is a transport exponent of the same type as the conductivity exponent $t$ in ordinary percolation, which is $1.303(8)$ according to \cite{wlm12}. 
In \cref{eqah81}, $p^+\to (1-p_c)^+$ means $p^+$ approaches $1-p_c$ from above and $p^+\to (p_c)^-$ means $p^+$ approaches $p_c$ from below. 
By using finite size scaling analysis, we have that
\begin{equation}
\label{eqah82}
M\sim L^{-t'/\nu}
\end{equation}
where $\nu$ is the correlation length exponent in percolation, which is known to be $4/3$ \cite{n79}.

To investigate the relation given in \cref{eqah82}, we set $p^+=0.5927\approx p_c$ and network-dimensions $L\times L$ for $L$ between $50$ to $90$~links. 
The lowest numerically feasible $|\Delta P|$ are used in order to stay in the lower linear regime in \cref{eqah80}, specifically $2.8$~Pa/link~$\lessapprox |\Delta P|/L \lessapprox 5.8$~Pa/link. 
When operating at low $|\Delta P|$, the flow, which is mainly through the connected paths, stabilizes quickly and retains approximately a constant value compared to the fluctuating flow at higher $|\Delta P|$.
For these simulations, the flow is driven for approximately $40$~pore-volumes of fluid through the network, where one pore-volume is equal to the total volume of pore space in the network. 
The values of $Q$ are calculated by averaging over the last $20$~pore-volumes simulated.
Variation in the connected paths a network can have is covered by averaging the results over $50$~network realizations.
The results are shown in \cref{fig:logL_logM}, where we get $t'/\nu= 4.3\pm 1.0$, giving
\begin{equation}
    \label{eqah83}
t' = 5.7\pm 1.3\;.
\end{equation}

This is a huge value.  A possible explanation for the observed value is that the system is {\it not\/} at a critical point in spite of the geometry of the easy links and the connected paths indicating this.  In our argumentation, we have not taken into account the {\it empty\/} links, i.e., those links that do not contain any interfaces.  They will be indistinguishable from the easy links with respect to the dynamics. These empty links drive the system away from the percolation critical point, and Fig.\ \ref{fig:logL_logM} is in reality indicative of non-algebraic behavior. We have indicated this possible shift in transition in the phase diagram shown in Fig.\ \ref{fig1}.  

\section{Non-Darcy behavior}%
\label{non-darcy}

In the simulations done for this section, networks have dimensions $100\times 100$~links$^2$. For each $|\Delta P|$, the flow is driven for approximately $100$~pore-volumes of fluid through the network. This ensures the steady-state flow and the value of $Q$ in steady state is calculated by averaging over the total flow rate during approximately the last $25$~pore-volumes simulated.

\subsection{Hysteresis}%
\label{sub:hysteresis}

\begin{figure}[ht]
  \centering
  \includegraphics[width=1.0\linewidth]{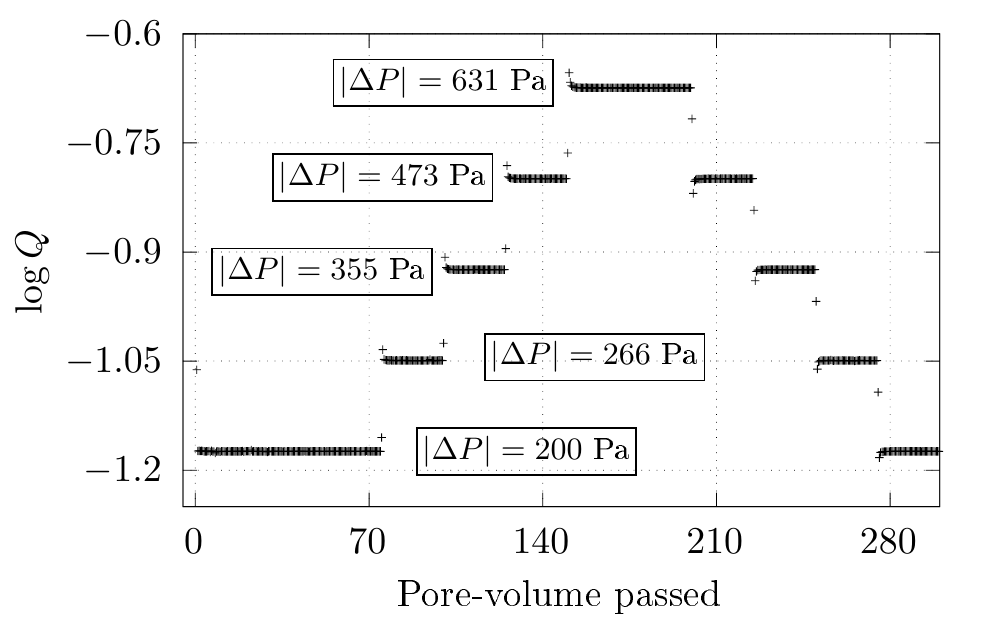}
  \caption{Increasing global pressure difference $|\Delta P|$ with the injected pore-volumes raises the volumetric flow rate $Q$ and subsequently decreasing $|\Delta P|$ returns $Q$ to the original value. $Q$ were measured in units ${\rm mm}^3/{\rm s}$.}%
  \label{fig:hysteresis}
\end{figure}

We pose here the question of whether there are any hysteretic effects from raising and lowering the pressure drop $|\Delta P|$ on the volumetric flow rate $Q$. The result is shown in \cref{fig:hysteresis}. 
With the passing of time, measuring in terms of injected pore-volumes, $|\Delta P|$ applied across a network is raised and then lowered in steps.
The $|\Delta P|$ values used, $200$~Pa, $266$~Pa, $355$~Pa, $473$~Pa and $631$~Pa, are from the lowest numerically feasible range.
It can be observed from \cref{fig:hysteresis} that whenever $|\Delta P|$ is returned to the same value, $Q$ also quickly stabilizes back to the previous value it had with the same $|\Delta P|$.
This shows that the steady state results generated using the DPN model do not depend on long-term memory \cite{esthfm13}.

\subsection{Volumetric flow rate dependence on pressure drop}%
\label{sub:res_volumetric_flow_rate_dependence_on_pressure}

\begin{figure}[ht]
  \centering
  \includegraphics[width=1.0\linewidth]{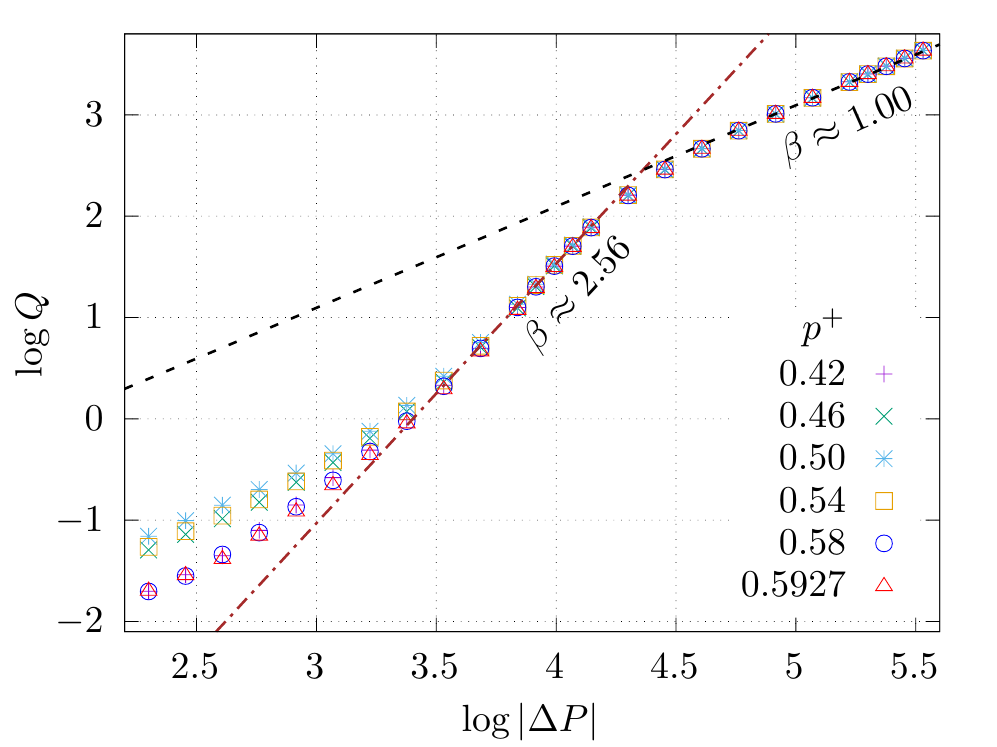}
  \caption{Total volumetric flow rate $Q$ as a function of global pressure difference $|\Delta P|$ in systems with different non-wetting grain occupation probabilities $p^+$. 
    The results of linear fit with slopes $\beta$ are included in the plot, where $\beta$ is the exponent in $Q\propto |\Delta P|^\beta$. 
    $Q$ were measured in units mm$^3$/s and $|\Delta P|$ were measured in units Pa. This figure is the basis for Eq.\ (\ref{eqah80}).}%
  \label{fig:p_DeltaP_Q}
\end{figure}

\begin{figure}[ht]
  \centering
  \includegraphics[width=0.95\linewidth]{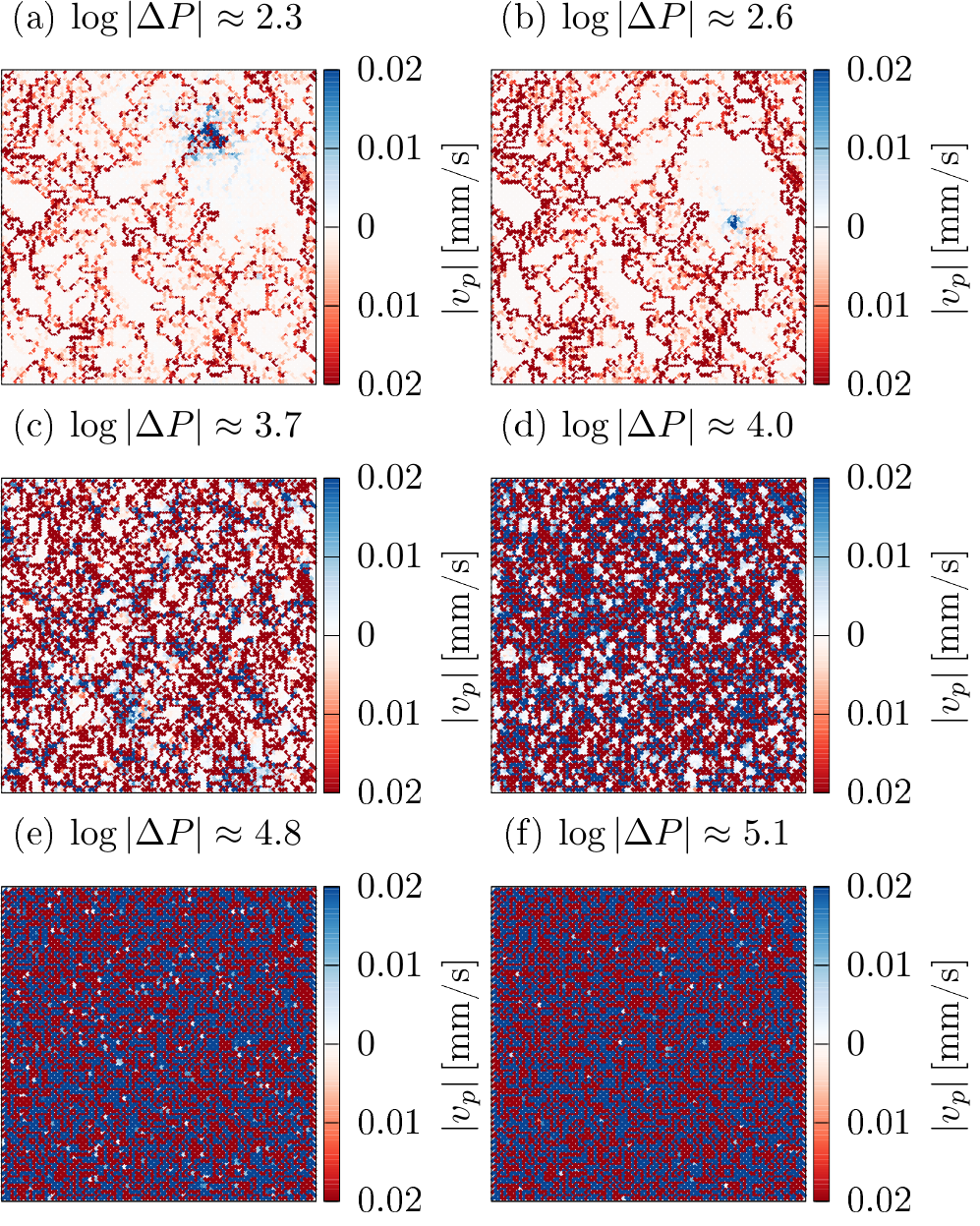}
  \caption{Maps of steady state averaged absolute velocities $|v_p|$ at different global pressure differences $|\Delta P|$, where velocities through links with wetting angle $\theta =90^\circ$ have shades of red and those through links with $\theta \in \{0^\circ,180^\circ\}$ have shades of blue. The network had non-wetting grain occupation probability $p^+=0.5$. $|\Delta P|$ were measured in units Pa.}%
  \label{fig:vp_map}
\end{figure}

The results relating $Q$ and $|\Delta P|$ in systems with zero $P_t$ and different values of $p^+$ are shown in \cref{fig:p_DeltaP_Q}.
We used $p^+ \in \{0.42,0.46,0.50,0.54,0.58,0.5927\}$ for the simulations.
For each of these $p^+$, the results were averaged over ten randomly chosen networks that have connected paths, meaning ten networks were randomly chosen from a subset of networks with zero threshold pressure $P_t$.
To assist the understanding of \cref{fig:p_DeltaP_Q}, velocity maps of a network with $p^+ = 0.5$ at various $|\Delta P|$ have been plotted in \cref{fig:vp_map}. 
The velocity maps show the steady state averaged absolute velocities, in other words, they show the average speed of the fluid.
The velocities are color coded so that those through neutral links with $\theta =90^\circ$ are in shades of red and the rest that are through links with $\theta \in \{0^\circ,180^\circ\}$ are in shades of blue. 
The results in \cref{fig:p_DeltaP_Q,fig:vp_map} show three regimes in terms of $\beta$ as indicated in Eq.\ (\ref{eqah80}). 

\begin{figure}[ht]
  \centering
  \includegraphics[width=0.95\linewidth]{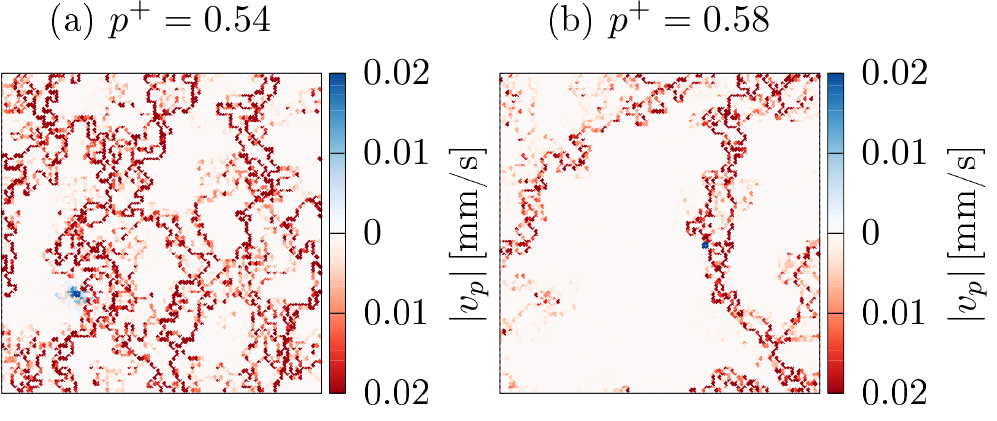}
  \caption{Maps of steady state averaged absolute velocities $|v_p|$ at $\log|\Delta P| \approx 2.3$ where $|\Delta P|$ is the global pressure difference. The velocities through links with wetting angle $\theta =90^\circ$ have shades of red and those through links with $\theta \in \{0^\circ,180^\circ\}$ have shades of blue. $p^+$ is the non-wetting grain occupation probability.}%
  \label{fig:vp_map5458}
\end{figure}

The lowest regime in \cref{eqah80} seems to correspond to $\log|\Delta P| \lessapprox 2.8$ in \cref{fig:p_DeltaP_Q}, in other words $|\Delta P| /L \lessapprox 6.3$~Pa/link.
The transition between this regime to the next is more gradual for $p^+$ away from $0.5$ in \cref{fig:p_DeltaP_Q}. 
In this regime with very low $|\Delta P|$, we find $\beta = 1.00 \pm 0.01$.
The velocity maps of a network with $p^+ = 0.5$ at two different $|\Delta P|$ in this regime are shown in \cref{fig:vp_map}(a)~and~(b) and they indicate that the flow is mainly through the neutrally wet (red) links.
When increasing $\log|\Delta P|$ from \cref{fig:vp_map}(a) to \cref{fig:vp_map}(b), the impact mainly manifests in the increase of the speed of the fluids rather than the creation of new paths. Therefore, it makes sense that the flow remains Darcy-like with $\beta$ approximately equal to $1$.  In the lowest regime in \cref{fig:p_DeltaP_Q}, it is apparent that the mobility $M$ in \cref{eqah80} decreases when $p^+$ moves away from $0.5$ towards $p_c\approx 0.5927$ and $1-p_c\approx 0.4073$.  In this regime, the flow is mainly through the connected paths.  
The network has more connected-path links, and transports more fluid for the same $|\Delta P|$ hence resulting in a larger $Q$, meaning a larger $M$ when $p^+$ moves towards 0.5. 
For instance, at $\log|\Delta P| \approx 2.3$, the number of active connected-path links is high at $p^+=0.5$, as can be seen in \cref{fig:vp_map}(a), slightly lower at $p^+ =0.54$, as can be seen in \cref{fig:vp_map5458}(a), and significantly lower at $p^+ =0.58$, as can be seen in \cref{fig:vp_map5458}(b), making $Q$ at $p^+ =0.58$ significantly less than in the other two cases.

The middle regime in \cref{eqah80} seems to corresponds to $3.3 \lessapprox \log|\Delta P| \lessapprox 4.1$ in \cref{fig:p_DeltaP_Q}, in other words $20.0$~Pa/Link~$\lessapprox |\Delta P|/L \lessapprox 125.9$~Pa/link.
Here, the exponent in \cref{eqah80} is $\beta =\beta_3= 2.56 \pm 0.05$ and $M_m$ is the same for all $p^+$ examined.
When $\log|\Delta P|$ increases from \cref{fig:vp_map}(c) to \cref{fig:vp_map}(d) in this regime, the velocity maps show that there is a significant increase in the number of flow carrying links, meaning $Q$ increases significantly also.
The opening of new paths in addition to increased flow in the already active paths explains $\beta$ being large. 
At this level of $|\Delta P|$, $M_m$ and $\beta$ being the same for all $p^+$ examined makes sense as the connected paths that differentiate networks with different $p^+$ no longer are the main contributors to the flow.  

The highest regime in \cref{eqah80} seems to corresponds to $\log|\Delta P| \gtrapprox 4.5$ in \cref{fig:p_DeltaP_Q}, in other words $|\Delta P|/L \gtrapprox 316.2$~Pa/link.
Here, the exponent in \cref{eqah80} is $\beta = 1.00 \pm 0.01$ and $M_D$ is the same for all $p^+$ examined.
The velocity maps taken from two different points in this regime are shown in \cref{fig:vp_map}(e)~and~(f).
In both cases, almost all the links in the network are carrying flow regardless of their wettability, hence increasing $|\Delta P|$ does not create new paths.
The effect of capillary barriers in the links becomes insignificant in comparison to the enormous pressure drop across the links, making all $p^+$ produce the same $Q$ at the same $|\Delta P|$. Increasing $|\Delta P|$ in this regime, increases $Q$ linearly, indicative of Darcy flow.

\begin{figure}[ht]
  \centering
  \includegraphics[width=1.0\linewidth]{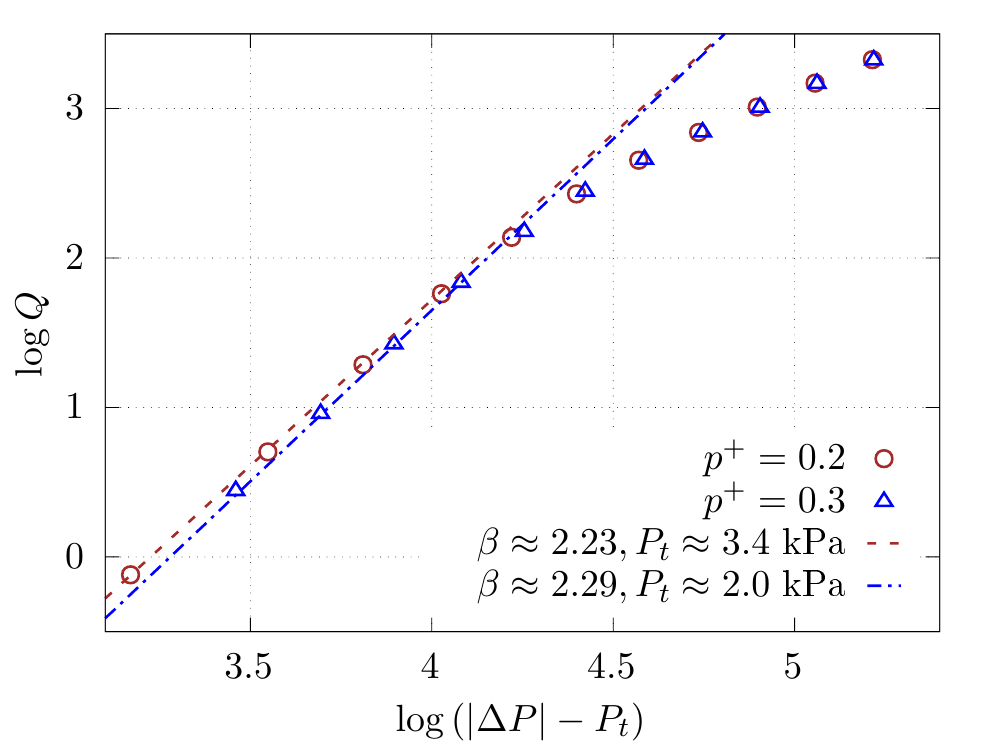}
  \caption{Total volumetric flow rate $Q$ as a function of effective pressure in systems with different non-wetting grain occupation probabilities $p^+$. Effective pressure is the difference between the global pressure difference $|\Delta P|$ and the threshold pressure $P_t$. 
    The results of linear fit with slopes $\beta$ are included in the plot, where $\beta$ is the exponent in $Q\propto \left(|\Delta P| - P_t\right)^\beta$. 
    $Q$ and $|\Delta P|$ were measured in the units of ${\rm mm}^3/{\rm s}$ and Pa respectively.}%
  \label{fig:p2n3_DeltaP_Q}
\end{figure}

As the results in \cref{easy} show, there are very few to zero connected paths outside of the range $1-p_c\le p^+\le p_c$ examined in \cref{fig:p_DeltaP_Q}.
If $p^+$ were very close to the range examined in \cref{fig:p_DeltaP_Q}, the behavior of $\beta$ and $M$ would have been expected to be the same as in \cref{fig:p_DeltaP_Q} since the flow will similarly be carried by the connected paths.
To test $p^+$ further away, simulations have been performed with $p^+ = 0.2$~and~$0.3$ and the results are shown in \cref{fig:p2n3_DeltaP_Q}.  
Here, $P_t$ is not zero, unlike the systems used for \cref{fig:p_DeltaP_Q} and corresponding constitutive \cref{eqah80}.  In this case, we find a constitutive equation 
\begin{equation}
    \label{eqah85}
    Q=\ \left\{  
    \begin{array}{ll}
         0 &  \mbox{if $|\Delta P| \le P_t$}\;,\\ 
        M_m(|\Delta P|-P_t)^\beta &  \mbox{if $P_t <|\Delta P| < P_u$}\;,\\
        M_D(|\Delta P|-P_t) &  \mbox{if $|\Delta P| < P_u$}\;,\\
    \end{array}                
    \right.  
\end{equation}
where $P_u$ is the crossover pressure between non-linear and Darcy behavior. 
By varying $P_t$ from $0.00$~Pa to the lowest $|\Delta P|$ in the data sets with an increment of $0.01$~Pa, mathematical linear fits with slopes $\beta$ were calculated at the lowest pressures to find the candidate that gave the least room-mean-square error.
This gave $\beta = 2.23 \pm 0.05$ and $P_t = \left(3.4 \pm 0.5\right)$~kPa for $p^+ = 0.2$ and $\beta = 2.29 \pm 0.05$ and $P_t = \left(2.0 \pm 0.5\right)$~kPa for $p^+ = 0.3$.
The regime these $\beta$ correspond to is the middle regime discussed in \cref{fig:p_DeltaP_Q} where the behavior was also non-linear due to the capillary barriers created by the interfaces between the two fluids. 
\textcite{fyhn2021rheology} has observed $\beta>1$ behavior even in networks with the same wetting angle everywhere which would be the same as having $p^+\to 0.0$~or~$1.0$ here. 
Due to the lack of connected paths in systems with $p^+ = 0.2$~and~$0.3$, the lowest regime in \cref{fig:p_DeltaP_Q} does not appear for the results in \cref{fig:p2n3_DeltaP_Q}.
Lastly, the highest regime where $\beta\approx 1$ should occur for all where flow pushes through almost the entire network and there is almost no influence of $p^+$.
That is indeed what we see in \cref{fig:p2n3_DeltaP_Q} also.

\section{Conclusion}%
\label{sec:conclusion}

We studied the effect of having porous media consisting of randomly mixed dual-wettability grains on the immiscible two-phase flow using a dynamic pore network model.
The model treats the interfacial tension between the two fluids similarly to a model introduced by Irannezhad et al.~\cite{ipbrz22,ipbrz23}. 
The model has two parameters, the saturation $S_w$ and the probability $p^+$ to have a grain of ``$+$'' type. The model, which is explained in Fig.\ \ref{fig:DPN_sketch}, contains links (pores) of three types when filled with two immiscible fluids A and B: Links that are wetting with respect to fluid type A, links that are wetting with respect to fluid type B and {\it easy links\/} where there are no capillary forces associated with interfaces.  The parameter $p^+$ controls the number of links generating capillary forces and easy links.  The model has a rich phase diagram, sketched in Fig.\ \ref{fig1}. There is a region $1-p_c\le p^+ \le p_c$, where $p_c$ is the site percolation threshold, where the easy links form connected paths across the network.  Outside this region, i.e., for $p^+\ll 1-p_c$ or $p^+\gg p_c$, easy links do not percolate.  We find two classes of constitutive equations for volumetric flow rate $Q$ vs.\ pressure drop $|\Delta P|$: For $1-p_c\le p^+ \le p_c$ we observed a constitutive equation as in Eq.~(\ref{eqah80}), see Fig.\ \ref{fig:p_DeltaP_Q}, whereas for $p^+\ll 1-p_c$ or $p^+\gg p_c$, we observed a constitutive equation (\ref{eqah85}), see Fig.\ \ref{fig:p2n3_DeltaP_Q}. The crucial point that distinguishes these is whether there is a non-zero threshold pressure $P_t$.

When $1-p_c\le p^+ \le p_c$ we observed the following:
At the regimes with lowest and highest $|\Delta P|$, it seems that $\beta = 1.00 \pm 0.01$, because there is no significant change in the paths fluids are flowing through and increasing $|\Delta P|$ only increases the flow in the already active paths.
At the lowest $|\Delta P|$, the flow is mainly always through connected paths with zero resistance.
When $p^+\to 0.5$ in this regime, there are more connected paths which means more fluid gets transported, making $Q$ hence $M$ higher.  
At the highest $|\Delta P|$, almost the entire network is always active.
On the other hand, $\beta > 1$ in the middle regime where an increase in $|\Delta P|$ increases the flow in the active paths and in addition opens new conducting paths.
In the middle and the highest regimes, the flow is no longer mainly through the connected paths and the differences between the pressures across the links and the capillary barriers in the links are large.
With the diminished role of the connected paths and capillary barriers at higher pressure drops, $M_m$ and $M_D$ do not depend strongly on $p^+$.
The exponent in the middle regime was found to be $\beta = \beta_3= 2.56 \pm 0.05$. We saw no systematic dependence of $\beta$ on $p^+$.

For $p^+ = 0.2$, however, $\beta = 2.23 \pm 0.05$ and $P_t = \left(3.4 \pm 0.5\right)$~kPa, and for $p^+ = 0.3$, $\beta = 2.29 \pm 0.05$ and $P_t = \left(2.0 \pm 0.5\right)$~kPa.  Due to the necessity of determining $P_t$ and $\beta$ simultaneously at these $p^+$ values, there is more uncertainty associated with the measurements of $\beta$. It is not possible to verify or falsify whether there is a fixed $\beta=\beta_2$, or whether it depends on $p^+$ and $S_w$.  

The existence of connecting paths is a percolation problem. They disappear when $p^+ \to (1-p_c)^+$ or $p^+ \to (p_c)^-$.  It would therefore be expected that the mobility $M$ defined in Eq.\ (\ref{eqah80}) would exhibit a critical behavior similar to the conductance near a critical point.  By making the hypothesis that $M$ behaves as in Eq.\ (\ref{eqah81}) and using finite size scaling, we determined $t' \approx 5.7$, see
Fig.\ \ref{fig:logL_logM}.  This is a huge value and raises the suspicion that the system is {\it not\/} critical where percolation theory dictates that it should be. Possible suspects for causing this push away from criticality are the links that do not contain interfaces.  They are not easy links, but they have precisely the same effect on the dynamics of the flow as the easy links.  If this is so, the transition lines would then be shifted as shown in Fig.\ \ref{fig1}. 

We have only explored a small part of the phase diagram of this rich model in this first study.  The phase diagram should be investigated in more detail and over a wider range of parameters.  The nature of the transition lines is  as of now unknown, and should also be further investigated. 
There are percolation transitions in the model.  Where are they and what are their properties as the transport is not through percolation clusters? 


\section{Declaration}%
\label{sec:declaration}

\noindent \textbf{Author Contributions}:
HF performed the numerical simulations and the data analysis and wrote the first draft of the manuscript.  HF wrote the code specific for this project based on algorithms written by SS. AH and SS suggested the idea of the problem. AH worked out the relation to percolation theory.

\noindent \textbf{Funding}: 
This work was supported by the Research Council of Norway through its Center of Excellence funding scheme, project number 262644.

\noindent \textbf{Acknowledgment}: 
We thank S.\ B.\ Santra and D.\ Jnana for discussions on the percolation problem that this system poses.





\bibliography{bibliography.bib}

\begin{thebibliography}{48}%
\makeatletter
\providecommand \@ifxundefined [1]{%
 \@ifx{#1\undefined}
}%
\providecommand \@ifnum [1]{%
 \ifnum #1\expandafter \@firstoftwo
 \else \expandafter \@secondoftwo
 \fi
}%
\providecommand \@ifx [1]{%
 \ifx #1\expandafter \@firstoftwo
 \else \expandafter \@secondoftwo
 \fi
}%
\providecommand \natexlab [1]{#1}%
\providecommand \enquote  [1]{``#1''}%
\providecommand \bibnamefont  [1]{#1}%
\providecommand \bibfnamefont [1]{#1}%
\providecommand \citenamefont [1]{#1}%
\providecommand \href@noop [0]{\@secondoftwo}%
\providecommand \href [0]{\begingroup \@sanitize@url \@href}%
\providecommand \@href[1]{\@@startlink{#1}\@@href}%
\providecommand \@@href[1]{\endgroup#1\@@endlink}%
\providecommand \@sanitize@url [0]{\catcode `\\12\catcode `\$12\catcode
  `\&12\catcode `\#12\catcode `\^12\catcode `\_12\catcode `\%12\relax}%
\providecommand \@@startlink[1]{}%
\providecommand \@@endlink[0]{}%
\providecommand \url  [0]{\begingroup\@sanitize@url \@url }%
\providecommand \@url [1]{\endgroup\@href {#1}{\urlprefix }}%
\providecommand \urlprefix  [0]{URL }%
\providecommand \Eprint [0]{\href }%
\providecommand \doibase [0]{http://dx.doi.org/}%
\providecommand \selectlanguage [0]{\@gobble}%
\providecommand \bibinfo  [0]{\@secondoftwo}%
\providecommand \bibfield  [0]{\@secondoftwo}%
\providecommand \translation [1]{[#1]}%
\providecommand \BibitemOpen [0]{}%
\providecommand \bibitemStop [0]{}%
\providecommand \bibitemNoStop [0]{.\EOS\space}%
\providecommand \EOS [0]{\spacefactor3000\relax}%
\providecommand \BibitemShut  [1]{\csname bibitem#1\endcsname}%
\let\auto@bib@innerbib\@empty
\bibitem [{\citenamefont {Ohm}(1827)}]{o27}%
  \BibitemOpen
  \bibfield  {author} {\bibinfo {author} {\bibfnamefont {G.~S.}\ \bibnamefont
  {Ohm}},\ }\href@noop {} {\emph {\bibinfo {title} {Die Galvanische Kette,
  matematisch bearbeitet}}}\ (\bibinfo  {publisher} {T. H. Riemann},\ \bibinfo
  {year} {1827})\BibitemShut {NoStop}%
\bibitem [{\citenamefont {Darcy}(1856)}]{d56}%
  \BibitemOpen
  \bibfield  {author} {\bibinfo {author} {\bibfnamefont {H.}~\bibnamefont
  {Darcy}},\ }\href@noop {} {\emph {\bibinfo {title} {Les fontaines publiques
  de la ville de Dijon}}}\ (\bibinfo  {publisher} {Dalmont},\ \bibinfo {year}
  {1856})\BibitemShut {NoStop}%
\bibitem [{\citenamefont {Whitaker}(1986)}]{w86}%
  \BibitemOpen
  \bibfield  {author} {\bibinfo {author} {\bibfnamefont {S.}~\bibnamefont
  {Whitaker}},\ }\href {\doibase 10.1007/BF01036523} {\bibfield  {journal}
  {\bibinfo  {journal} {Transport in porous media}\ }\textbf {\bibinfo {volume}
  {1}},\ \bibinfo {pages} {3} (\bibinfo {year} {1986})}\BibitemShut {NoStop}%
\bibitem [{\citenamefont {Wyckoff}\ and\ \citenamefont {Botset}(1936)}]{wb36}%
  \BibitemOpen
  \bibfield  {author} {\bibinfo {author} {\bibfnamefont {R.}~\bibnamefont
  {Wyckoff}}\ and\ \bibinfo {author} {\bibfnamefont {H.}~\bibnamefont
  {Botset}},\ }\href {\doibase 10.1063/1.1745402} {\bibfield  {journal}
  {\bibinfo  {journal} {Physics}\ }\textbf {\bibinfo {volume} {7}},\ \bibinfo
  {pages} {325} (\bibinfo {year} {1936})}\BibitemShut {NoStop}%
\bibitem [{\citenamefont {Barenblatt}\ \emph {et~al.}(2002)\citenamefont
  {Barenblatt}, \citenamefont {Patzek},\ and\ \citenamefont {Silin}}]{bps02}%
  \BibitemOpen
  \bibfield  {author} {\bibinfo {author} {\bibfnamefont {G.}~\bibnamefont
  {Barenblatt}}, \bibinfo {author} {\bibfnamefont {T.}~\bibnamefont {Patzek}},
  \ and\ \bibinfo {author} {\bibfnamefont {D.}~\bibnamefont {Silin}},\ }in\
  \href {\doibase 10.2118/75169-MS} {\emph {\bibinfo {booktitle} {Spe/doe
  improved oil recovery symposium}}}\ (\bibinfo {organization} {OnePetro},\
  \bibinfo {year} {2002})\BibitemShut {NoStop}%
\bibitem [{\citenamefont {Tallakstad}\ \emph
  {et~al.}(2009{\natexlab{a}})\citenamefont {Tallakstad}, \citenamefont
  {Knudsen}, \citenamefont {Ramstad}, \citenamefont {L{\o}voll}, \citenamefont
  {M{\aa}l{\o}y}, \citenamefont {Toussaint},\ and\ \citenamefont
  {Flekk{\o}y}}]{tallakstad2009steady}%
  \BibitemOpen
  \bibfield  {author} {\bibinfo {author} {\bibfnamefont {K.~T.}\ \bibnamefont
  {Tallakstad}}, \bibinfo {author} {\bibfnamefont {H.~A.}\ \bibnamefont
  {Knudsen}}, \bibinfo {author} {\bibfnamefont {T.}~\bibnamefont {Ramstad}},
  \bibinfo {author} {\bibfnamefont {G.}~\bibnamefont {L{\o}voll}}, \bibinfo
  {author} {\bibfnamefont {K.~J.}\ \bibnamefont {M{\aa}l{\o}y}}, \bibinfo
  {author} {\bibfnamefont {R.}~\bibnamefont {Toussaint}}, \ and\ \bibinfo
  {author} {\bibfnamefont {E.~G.}\ \bibnamefont {Flekk{\o}y}},\ }\href
  {\doibase https://doi.org/10.1103/PhysRevLett.102.074502} {\bibfield
  {journal} {\bibinfo  {journal} {Physical review letters}\ }\textbf {\bibinfo
  {volume} {102}},\ \bibinfo {pages} {074502} (\bibinfo {year}
  {2009}{\natexlab{a}})}\BibitemShut {NoStop}%
\bibitem [{\citenamefont {Tallakstad}\ \emph
  {et~al.}(2009{\natexlab{b}})\citenamefont {Tallakstad}, \citenamefont
  {L{\o}voll}, \citenamefont {Knudsen}, \citenamefont {Ramstad}, \citenamefont
  {Flekk{\o}y},\ and\ \citenamefont {M{\aa}l{\o}y}}]{tallakstad2009steadyE}%
  \BibitemOpen
  \bibfield  {author} {\bibinfo {author} {\bibfnamefont {K.~T.}\ \bibnamefont
  {Tallakstad}}, \bibinfo {author} {\bibfnamefont {G.}~\bibnamefont
  {L{\o}voll}}, \bibinfo {author} {\bibfnamefont {H.~A.}\ \bibnamefont
  {Knudsen}}, \bibinfo {author} {\bibfnamefont {T.}~\bibnamefont {Ramstad}},
  \bibinfo {author} {\bibfnamefont {E.~G.}\ \bibnamefont {Flekk{\o}y}}, \ and\
  \bibinfo {author} {\bibfnamefont {K.~J.}\ \bibnamefont {M{\aa}l{\o}y}},\
  }\href {\doibase https://doi.org/10.1103/PhysRevE.80.036308} {\bibfield
  {journal} {\bibinfo  {journal} {Physical Review E}\ }\textbf {\bibinfo
  {volume} {80}},\ \bibinfo {pages} {036308} (\bibinfo {year}
  {2009}{\natexlab{b}})}\BibitemShut {NoStop}%
\bibitem [{\citenamefont {Rassi}\ \emph {et~al.}(2011)\citenamefont {Rassi},
  \citenamefont {Codd},\ and\ \citenamefont {Seymour}}]{rcs11}%
  \BibitemOpen
  \bibfield  {author} {\bibinfo {author} {\bibfnamefont {E.~M.}\ \bibnamefont
  {Rassi}}, \bibinfo {author} {\bibfnamefont {S.~L.}\ \bibnamefont {Codd}}, \
  and\ \bibinfo {author} {\bibfnamefont {J.~D.}\ \bibnamefont {Seymour}},\
  }\href {\doibase 10.1088/1367-2630/13/1/015007} {\bibfield  {journal}
  {\bibinfo  {journal} {New Journal of Physics}\ }\textbf {\bibinfo {volume}
  {13}},\ \bibinfo {pages} {015007} (\bibinfo {year} {2011})}\BibitemShut
  {NoStop}%
\bibitem [{\citenamefont {Aursj{\o}}\ \emph {et~al.}(2014)\citenamefont
  {Aursj{\o}}, \citenamefont {Erpelding}, \citenamefont {Tallakstad},
  \citenamefont {Flekk{\o}y}, \citenamefont {Hansen},\ and\ \citenamefont
  {M{\aa}l{\o}y}}]{aetfhm14}%
  \BibitemOpen
  \bibfield  {author} {\bibinfo {author} {\bibfnamefont {O.}~\bibnamefont
  {Aursj{\o}}}, \bibinfo {author} {\bibfnamefont {M.}~\bibnamefont
  {Erpelding}}, \bibinfo {author} {\bibfnamefont {K.~T.}\ \bibnamefont
  {Tallakstad}}, \bibinfo {author} {\bibfnamefont {E.~G.}\ \bibnamefont
  {Flekk{\o}y}}, \bibinfo {author} {\bibfnamefont {A.}~\bibnamefont {Hansen}},
  \ and\ \bibinfo {author} {\bibfnamefont {K.~J.}\ \bibnamefont
  {M{\aa}l{\o}y}},\ }\href {\doibase 10.3389/fphy.2014.00063} {\bibfield
  {journal} {\bibinfo  {journal} {Frontiers in physics}\ }\textbf {\bibinfo
  {volume} {2}},\ \bibinfo {pages} {63} (\bibinfo {year} {2014})}\BibitemShut
  {NoStop}%
\bibitem [{\citenamefont {Sinha}\ \emph {et~al.}(2017)\citenamefont {Sinha},
  \citenamefont {Bender}, \citenamefont {Danczyk}, \citenamefont {Keepseagle},
  \citenamefont {Prather}, \citenamefont {Bray}, \citenamefont {Thrane},
  \citenamefont {Seymour}, \citenamefont {Codd},\ and\ \citenamefont
  {Hansen}}]{sh17}%
  \BibitemOpen
  \bibfield  {author} {\bibinfo {author} {\bibfnamefont {S.}~\bibnamefont
  {Sinha}}, \bibinfo {author} {\bibfnamefont {A.~T.}\ \bibnamefont {Bender}},
  \bibinfo {author} {\bibfnamefont {M.}~\bibnamefont {Danczyk}}, \bibinfo
  {author} {\bibfnamefont {K.}~\bibnamefont {Keepseagle}}, \bibinfo {author}
  {\bibfnamefont {C.~A.}\ \bibnamefont {Prather}}, \bibinfo {author}
  {\bibfnamefont {J.~M.}\ \bibnamefont {Bray}}, \bibinfo {author}
  {\bibfnamefont {L.~W.}\ \bibnamefont {Thrane}}, \bibinfo {author}
  {\bibfnamefont {J.~D.}\ \bibnamefont {Seymour}}, \bibinfo {author}
  {\bibfnamefont {S.~L.}\ \bibnamefont {Codd}}, \ and\ \bibinfo {author}
  {\bibfnamefont {A.}~\bibnamefont {Hansen}},\ }\href {\doibase
  10.1007/s11242-017-0874-4} {\bibfield  {journal} {\bibinfo  {journal}
  {Transport in porous media}\ }\textbf {\bibinfo {volume} {119}},\ \bibinfo
  {pages} {77} (\bibinfo {year} {2017})}\BibitemShut {NoStop}%
\bibitem [{\citenamefont {Gao}\ \emph {et~al.}(2020)\citenamefont {Gao},
  \citenamefont {Lin}, \citenamefont {Bijeljic},\ and\ \citenamefont
  {Blunt}}]{glbb20}%
  \BibitemOpen
  \bibfield  {author} {\bibinfo {author} {\bibfnamefont {Y.}~\bibnamefont
  {Gao}}, \bibinfo {author} {\bibfnamefont {Q.}~\bibnamefont {Lin}}, \bibinfo
  {author} {\bibfnamefont {B.}~\bibnamefont {Bijeljic}}, \ and\ \bibinfo
  {author} {\bibfnamefont {M.~J.}\ \bibnamefont {Blunt}},\ }\href {\doibase
  10.1103/PhysRevFluids.5.013801} {\bibfield  {journal} {\bibinfo  {journal}
  {Physical Review Fluids}\ }\textbf {\bibinfo {volume} {5}},\ \bibinfo {pages}
  {013801} (\bibinfo {year} {2020})}\BibitemShut {NoStop}%
\bibitem [{\citenamefont {Zhang}\ \emph {et~al.}(2021)\citenamefont {Zhang},
  \citenamefont {Bijeljic}, \citenamefont {Gao}, \citenamefont {Lin},\ and\
  \citenamefont {Blunt}}]{zbglb21}%
  \BibitemOpen
  \bibfield  {author} {\bibinfo {author} {\bibfnamefont {Y.}~\bibnamefont
  {Zhang}}, \bibinfo {author} {\bibfnamefont {B.}~\bibnamefont {Bijeljic}},
  \bibinfo {author} {\bibfnamefont {Y.}~\bibnamefont {Gao}}, \bibinfo {author}
  {\bibfnamefont {Q.}~\bibnamefont {Lin}}, \ and\ \bibinfo {author}
  {\bibfnamefont {M.~J.}\ \bibnamefont {Blunt}},\ }\href {\doibase
  10.1029/2020GL090477} {\bibfield  {journal} {\bibinfo  {journal} {Geophysical
  Research Letters}\ }\textbf {\bibinfo {volume} {48}},\ \bibinfo {pages}
  {e2020GL090477} (\bibinfo {year} {2021})}\BibitemShut {NoStop}%
\bibitem [{\citenamefont {Zhang}\ \emph {et~al.}(2022)\citenamefont {Zhang},
  \citenamefont {Bijeljic},\ and\ \citenamefont {Blunt}}]{zbb22}%
  \BibitemOpen
  \bibfield  {author} {\bibinfo {author} {\bibfnamefont {Y.}~\bibnamefont
  {Zhang}}, \bibinfo {author} {\bibfnamefont {B.}~\bibnamefont {Bijeljic}}, \
  and\ \bibinfo {author} {\bibfnamefont {M.~J.}\ \bibnamefont {Blunt}},\ }\href
  {\doibase 10.1017/jfm.2021.1148} {\bibfield  {journal} {\bibinfo  {journal}
  {Journal of Fluid Mechanics}\ }\textbf {\bibinfo {volume} {934}},\ \bibinfo
  {pages} {R3} (\bibinfo {year} {2022})}\BibitemShut {NoStop}%
\bibitem [{\citenamefont {Gr{\o}va}\ and\ \citenamefont {Hansen}(2011)}]{gh11}%
  \BibitemOpen
  \bibfield  {author} {\bibinfo {author} {\bibfnamefont {M.}~\bibnamefont
  {Gr{\o}va}}\ and\ \bibinfo {author} {\bibfnamefont {A.}~\bibnamefont
  {Hansen}},\ }in\ \href {\doibase 10.1088/1742-6596/319/1/012009} {\emph
  {\bibinfo {booktitle} {Journal of Physics: Conference Series}}},\ Vol.\
  \bibinfo {volume} {319}\ (\bibinfo {organization} {IOP Publishing},\ \bibinfo
  {year} {2011})\ p.\ \bibinfo {pages} {012009}\BibitemShut {NoStop}%
\bibitem [{\citenamefont {Sinha}\ and\ \citenamefont {Hansen}(2012)}]{sh12}%
  \BibitemOpen
  \bibfield  {author} {\bibinfo {author} {\bibfnamefont {S.}~\bibnamefont
  {Sinha}}\ and\ \bibinfo {author} {\bibfnamefont {A.}~\bibnamefont {Hansen}},\
  }\href {\doibase 10.1209/0295-5075/99/44004} {\bibfield  {journal} {\bibinfo
  {journal} {Europhysics Letters}\ }\textbf {\bibinfo {volume} {99}},\ \bibinfo
  {pages} {44004} (\bibinfo {year} {2012})}\BibitemShut {NoStop}%
\bibitem [{\citenamefont {Sinha}\ \emph {et~al.}(2013)\citenamefont {Sinha},
  \citenamefont {Hansen}, \citenamefont {Bedeaux},\ and\ \citenamefont
  {Kjelstrup}}]{shbk13}%
  \BibitemOpen
  \bibfield  {author} {\bibinfo {author} {\bibfnamefont {S.}~\bibnamefont
  {Sinha}}, \bibinfo {author} {\bibfnamefont {A.}~\bibnamefont {Hansen}},
  \bibinfo {author} {\bibfnamefont {D.}~\bibnamefont {Bedeaux}}, \ and\
  \bibinfo {author} {\bibfnamefont {S.}~\bibnamefont {Kjelstrup}},\ }\href
  {\doibase 10.1209/0295-5075/99/44004} {\bibfield  {journal} {\bibinfo
  {journal} {Physical Review E}\ }\textbf {\bibinfo {volume} {87}},\ \bibinfo
  {pages} {025001} (\bibinfo {year} {2013})}\BibitemShut {NoStop}%
\bibitem [{\citenamefont {Xu}\ and\ \citenamefont {Wang}(2014)}]{xw14}%
  \BibitemOpen
  \bibfield  {author} {\bibinfo {author} {\bibfnamefont {X.}~\bibnamefont
  {Xu}}\ and\ \bibinfo {author} {\bibfnamefont {X.}~\bibnamefont {Wang}},\
  }\href {\doibase 10.1103/PhysRevE.90.023010} {\bibfield  {journal} {\bibinfo
  {journal} {Physical Review E}\ }\textbf {\bibinfo {volume} {90}},\ \bibinfo
  {pages} {023010} (\bibinfo {year} {2014})}\BibitemShut {NoStop}%
\bibitem [{\citenamefont {Yiotis}\ \emph {et~al.}(2019)\citenamefont {Yiotis},
  \citenamefont {Dollari}, \citenamefont {Kainourgiakis}, \citenamefont
  {Salin},\ and\ \citenamefont {Talon}}]{ydkst19}%
  \BibitemOpen
  \bibfield  {author} {\bibinfo {author} {\bibfnamefont {A.}~\bibnamefont
  {Yiotis}}, \bibinfo {author} {\bibfnamefont {A.}~\bibnamefont {Dollari}},
  \bibinfo {author} {\bibfnamefont {M.}~\bibnamefont {Kainourgiakis}}, \bibinfo
  {author} {\bibfnamefont {D.}~\bibnamefont {Salin}}, \ and\ \bibinfo {author}
  {\bibfnamefont {L.}~\bibnamefont {Talon}},\ }\href {\doibase
  10.1103/PhysRevFluids.4.114302} {\bibfield  {journal} {\bibinfo  {journal}
  {Physical Review Fluids}\ }\textbf {\bibinfo {volume} {4}},\ \bibinfo {pages}
  {114302} (\bibinfo {year} {2019})}\BibitemShut {NoStop}%
\bibitem [{\citenamefont {Roy}\ \emph {et~al.}(2019{\natexlab{a}})\citenamefont
  {Roy}, \citenamefont {Hansen},\ and\ \citenamefont {Sinha}}]{rhs19}%
  \BibitemOpen
  \bibfield  {author} {\bibinfo {author} {\bibfnamefont {S.}~\bibnamefont
  {Roy}}, \bibinfo {author} {\bibfnamefont {A.}~\bibnamefont {Hansen}}, \ and\
  \bibinfo {author} {\bibfnamefont {S.}~\bibnamefont {Sinha}},\ }\href
  {\doibase doi.org/10.3389/fphy.2019.00092} {\bibfield  {journal} {\bibinfo
  {journal} {Frontiers in Physics}\ }\textbf {\bibinfo {volume} {7}},\ \bibinfo
  {pages} {92} (\bibinfo {year} {2019}{\natexlab{a}})}\BibitemShut {NoStop}%
\bibitem [{\citenamefont {Roy}\ \emph {et~al.}(2019{\natexlab{b}})\citenamefont
  {Roy}, \citenamefont {Sinha},\ and\ \citenamefont {Hansen}}]{rsh19}%
  \BibitemOpen
  \bibfield  {author} {\bibinfo {author} {\bibfnamefont {S.}~\bibnamefont
  {Roy}}, \bibinfo {author} {\bibfnamefont {S.}~\bibnamefont {Sinha}}, \ and\
  \bibinfo {author} {\bibfnamefont {A.}~\bibnamefont {Hansen}},\ }\href
  {\doibase 10.48550/arXiv.1912.05248} {\bibfield  {journal} {\bibinfo
  {journal} {arXiv preprint arXiv:1912.05248}\ } (\bibinfo {year}
  {2019}{\natexlab{b}}),\ 10.48550/arXiv.1912.05248}\BibitemShut {NoStop}%
\bibitem [{\citenamefont {Fyhn}\ \emph {et~al.}(2021)\citenamefont {Fyhn},
  \citenamefont {Sinha}, \citenamefont {Roy},\ and\ \citenamefont
  {Hansen}}]{fyhn2021rheology}%
  \BibitemOpen
  \bibfield  {author} {\bibinfo {author} {\bibfnamefont {H.}~\bibnamefont
  {Fyhn}}, \bibinfo {author} {\bibfnamefont {S.}~\bibnamefont {Sinha}},
  \bibinfo {author} {\bibfnamefont {S.}~\bibnamefont {Roy}}, \ and\ \bibinfo
  {author} {\bibfnamefont {A.}~\bibnamefont {Hansen}},\ }\href {\doibase
  https://doi.org/10.1007/s11242-021-01674-3} {\bibfield  {journal} {\bibinfo
  {journal} {Transport in Porous Media}\ }\textbf {\bibinfo {volume} {139}},\
  \bibinfo {pages} {491} (\bibinfo {year} {2021})}\BibitemShut {NoStop}%
\bibitem [{\citenamefont {Sales}\ \emph {et~al.}(2022)\citenamefont {Sales},
  \citenamefont {Seybold}, \citenamefont {Oliveira},\ and\ \citenamefont
  {Andrade}}]{ssoa22}%
  \BibitemOpen
  \bibfield  {author} {\bibinfo {author} {\bibfnamefont {J.}~\bibnamefont
  {Sales}}, \bibinfo {author} {\bibfnamefont {H.~J.}\ \bibnamefont {Seybold}},
  \bibinfo {author} {\bibfnamefont {C.~L.}\ \bibnamefont {Oliveira}}, \ and\
  \bibinfo {author} {\bibfnamefont {J.~S.}\ \bibnamefont {Andrade}},\ }\href
  {\doibase 10.3389/fphy.2022.860190} {\bibfield  {journal} {\bibinfo
  {journal} {Frontiers in Physics}\ ,\ \bibinfo {pages} {170}} (\bibinfo {year}
  {2022})}\BibitemShut {NoStop}%
\bibitem [{\citenamefont {Feder}\ \emph {et~al.}(2022)\citenamefont {Feder},
  \citenamefont {Flekk{\o}y},\ and\ \citenamefont {Hansen}}]{ffh22}%
  \BibitemOpen
  \bibfield  {author} {\bibinfo {author} {\bibfnamefont {J.}~\bibnamefont
  {Feder}}, \bibinfo {author} {\bibfnamefont {E.~G.}\ \bibnamefont
  {Flekk{\o}y}}, \ and\ \bibinfo {author} {\bibfnamefont {A.}~\bibnamefont
  {Hansen}},\ }\href {\doibase 10.1017/9781009100717} {\emph {\bibinfo {title}
  {Physics of Flow in Porous Media}}}\ (\bibinfo  {publisher} {Cambridge
  University Press},\ \bibinfo {year} {2022})\BibitemShut {NoStop}%
\bibitem [{\citenamefont {Lanza}\ \emph {et~al.}(2022)\citenamefont {Lanza},
  \citenamefont {Rosso}, \citenamefont {Talon},\ and\ \citenamefont
  {Hansen}}]{lhrt22}%
  \BibitemOpen
  \bibfield  {author} {\bibinfo {author} {\bibfnamefont {F.}~\bibnamefont
  {Lanza}}, \bibinfo {author} {\bibfnamefont {A.}~\bibnamefont {Rosso}},
  \bibinfo {author} {\bibfnamefont {L.}~\bibnamefont {Talon}}, \ and\ \bibinfo
  {author} {\bibfnamefont {A.}~\bibnamefont {Hansen}},\ }\href {\doibase
  10.1007/s11242-022-01848-7} {\bibfield  {journal} {\bibinfo  {journal}
  {Transport in Porous Media}\ }\textbf {\bibinfo {volume} {145}},\ \bibinfo
  {pages} {245} (\bibinfo {year} {2022})}\BibitemShut {NoStop}%
\bibitem [{\citenamefont {Cheon}\ \emph {et~al.}(2023)\citenamefont {Cheon},
  \citenamefont {Fyhn}, \citenamefont {Hansen}, \citenamefont {Wilhelmsen},\
  and\ \citenamefont {Sinha}}]{cfhws23}%
  \BibitemOpen
  \bibfield  {author} {\bibinfo {author} {\bibfnamefont {H.~L.}\ \bibnamefont
  {Cheon}}, \bibinfo {author} {\bibfnamefont {H.}~\bibnamefont {Fyhn}},
  \bibinfo {author} {\bibfnamefont {A.}~\bibnamefont {Hansen}}, \bibinfo
  {author} {\bibfnamefont {{\O}.}~\bibnamefont {Wilhelmsen}}, \ and\ \bibinfo
  {author} {\bibfnamefont {S.}~\bibnamefont {Sinha}},\ }\href {\doibase
  10.1007/s11242-022-01893-2} {\bibfield  {journal} {\bibinfo  {journal}
  {Transport in Porous Media}\ ,\ \bibinfo {pages} {1}} (\bibinfo {year}
  {2023})}\BibitemShut {NoStop}%
\bibitem [{\citenamefont {Wilkinson}(1986)}]{wilkinson86}%
  \BibitemOpen
  \bibfield  {author} {\bibinfo {author} {\bibfnamefont {D.}~\bibnamefont
  {Wilkinson}},\ }\href {\doibase 10.1103/PhysRevA.34.1380} {\bibfield
  {journal} {\bibinfo  {journal} {Phys. Rev. A}\ }\textbf {\bibinfo {volume}
  {34}},\ \bibinfo {pages} {1380} (\bibinfo {year} {1986})}\BibitemShut
  {NoStop}%
\bibitem [{\citenamefont {Stauffer}\ and\ \citenamefont
  {Aharony}(2018)}]{sa18}%
  \BibitemOpen
  \bibfield  {author} {\bibinfo {author} {\bibfnamefont {D.}~\bibnamefont
  {Stauffer}}\ and\ \bibinfo {author} {\bibfnamefont {A.}~\bibnamefont
  {Aharony}},\ }\href@noop {} {\emph {\bibinfo {title} {Introduction to
  percolation theory}}}\ (\bibinfo  {publisher} {Taylor \& Francis},\ \bibinfo
  {year} {2018})\BibitemShut {NoStop}%
\bibitem [{\citenamefont {Roux}\ and\ \citenamefont {Herrmann}(1987)}]{rh87}%
  \BibitemOpen
  \bibfield  {author} {\bibinfo {author} {\bibfnamefont {S.}~\bibnamefont
  {Roux}}\ and\ \bibinfo {author} {\bibfnamefont {H.}~\bibnamefont
  {Herrmann}},\ }\href {\doibase 10.1209/0295-5075/4/11/003} {\bibfield
  {journal} {\bibinfo  {journal} {EUROPHYSics letters}\ }\textbf {\bibinfo
  {volume} {4}},\ \bibinfo {pages} {1227} (\bibinfo {year} {1987})}\BibitemShut
  {NoStop}%
\bibitem [{\citenamefont {Scheidegger}(1953)}]{s53}%
  \BibitemOpen
  \bibfield  {author} {\bibinfo {author} {\bibfnamefont {A.~E.}\ \bibnamefont
  {Scheidegger}},\ }\href@noop {} {\enquote {\bibinfo {title} {Theoretical
  models of porous matter: Prod},}\ } (\bibinfo {year} {1953})\BibitemShut
  {NoStop}%
\bibitem [{\citenamefont {Scheidegger}(2020)}]{s20}%
  \BibitemOpen
  \bibfield  {author} {\bibinfo {author} {\bibfnamefont {A.~E.}\ \bibnamefont
  {Scheidegger}},\ }in\ \href@noop {} {\emph {\bibinfo {booktitle} {The Physics
  of Flow Through Porous Media (3rd Edition)}}}\ (\bibinfo  {publisher}
  {University of Toronto press},\ \bibinfo {year} {2020})\BibitemShut {NoStop}%
\bibitem [{\citenamefont {Irannezhad}\ \emph {et~al.}(2022)\citenamefont
  {Irannezhad}, \citenamefont {Primkulov}, \citenamefont {Juanes},\ and\
  \citenamefont {Zhao}}]{ipbrz22}%
  \BibitemOpen
  \bibfield  {author} {\bibinfo {author} {\bibfnamefont {A.}~\bibnamefont
  {Irannezhad}}, \bibinfo {author} {\bibfnamefont {B.~K.}\ \bibnamefont
  {Primkulov}}, \bibinfo {author} {\bibfnamefont {R.}~\bibnamefont {Juanes}}, \
  and\ \bibinfo {author} {\bibfnamefont {B.}~\bibnamefont {Zhao}},\ }\href
  {\doibase 10.48550/arXiv.2207.01592} {\ ,\ \bibinfo {pages}
  {arXiv:2207.01592} (\bibinfo {year} {2022})}\BibitemShut {NoStop}%
\bibitem [{\citenamefont {Irannezhad}\ \emph {et~al.}(2023)\citenamefont
  {Irannezhad}, \citenamefont {Primkulov}, \citenamefont {Juanes},\ and\
  \citenamefont {Zhao}}]{ipbrz23}%
  \BibitemOpen
  \bibfield  {author} {\bibinfo {author} {\bibfnamefont {A.}~\bibnamefont
  {Irannezhad}}, \bibinfo {author} {\bibfnamefont {B.~K.}\ \bibnamefont
  {Primkulov}}, \bibinfo {author} {\bibfnamefont {R.}~\bibnamefont {Juanes}}, \
  and\ \bibinfo {author} {\bibfnamefont {B.}~\bibnamefont {Zhao}},\ }\href
  {\doibase 10.48550/arXiv.2302.03072} {\ ,\ \bibinfo {pages}
  {arXiv:2302.03072} (\bibinfo {year} {2023})}\BibitemShut {NoStop}%
\bibitem [{\citenamefont {Knudsen}\ and\ \citenamefont {Hansen}(2006)}]{kh06}%
  \BibitemOpen
  \bibfield  {author} {\bibinfo {author} {\bibfnamefont {H.~A.}\ \bibnamefont
  {Knudsen}}\ and\ \bibinfo {author} {\bibfnamefont {A.}~\bibnamefont
  {Hansen}},\ }\href {\doibase 10.1140/epjb/e2006-00019-y} {\bibfield
  {journal} {\bibinfo  {journal} {The European Physical Journal B-Condensed
  Matter and Complex Systems}\ }\textbf {\bibinfo {volume} {49}},\ \bibinfo
  {pages} {109} (\bibinfo {year} {2006})}\BibitemShut {NoStop}%
\bibitem [{\citenamefont {Grossman}\ and\ \citenamefont
  {Aharony}(1986)}]{ga86}%
  \BibitemOpen
  \bibfield  {author} {\bibinfo {author} {\bibfnamefont {T.}~\bibnamefont
  {Grossman}}\ and\ \bibinfo {author} {\bibfnamefont {A.}~\bibnamefont
  {Aharony}},\ }\href {\doibase 10.1088/0305-4470/19/12/009} {\bibfield
  {journal} {\bibinfo  {journal} {Journal of Physics A: Mathematical and
  General}\ }\textbf {\bibinfo {volume} {19}},\ \bibinfo {pages} {L745}
  (\bibinfo {year} {1986})}\BibitemShut {NoStop}%
\bibitem [{\citenamefont {Redner}(2007)}]{r07}%
  \BibitemOpen
  \bibfield  {author} {\bibinfo {author} {\bibfnamefont {S.}~\bibnamefont
  {Redner}},\ }\href {\doibase 10.48550/arXiv.0710.1105} {\bibfield  {journal}
  {\bibinfo  {journal} {arXiv preprint arXiv:0710.1105}\ } (\bibinfo {year}
  {2007}),\ 10.48550/arXiv.0710.1105}\BibitemShut {NoStop}%
\bibitem [{\citenamefont {Sinha}\ \emph {et~al.}(2021)\citenamefont {Sinha},
  \citenamefont {Gjennestad}, \citenamefont {Vassvik},\ and\ \citenamefont
  {Hansen}}]{sinha2019dynamic}%
  \BibitemOpen
  \bibfield  {author} {\bibinfo {author} {\bibfnamefont {S.}~\bibnamefont
  {Sinha}}, \bibinfo {author} {\bibfnamefont {M.~A.}\ \bibnamefont
  {Gjennestad}}, \bibinfo {author} {\bibfnamefont {M.}~\bibnamefont {Vassvik}},
  \ and\ \bibinfo {author} {\bibfnamefont {A.}~\bibnamefont {Hansen}},\ }\href
  {\doibase https://doi.org/10.3389/fphy.2020.548497} {\bibfield  {journal}
  {\bibinfo  {journal} {Frontiers in Physics}\ }\textbf {\bibinfo {volume}
  {8}},\ \bibinfo {pages} {567} (\bibinfo {year} {2021})}\BibitemShut {NoStop}%
\bibitem [{\citenamefont {Aker}\ \emph {et~al.}(1998)\citenamefont {Aker},
  \citenamefont {M{\aa}l{\o}y}, \citenamefont {Hansen},\ and\ \citenamefont
  {Batrouni}}]{amhb98}%
  \BibitemOpen
  \bibfield  {author} {\bibinfo {author} {\bibfnamefont {E.}~\bibnamefont
  {Aker}}, \bibinfo {author} {\bibfnamefont {K.~J.}\ \bibnamefont
  {M{\aa}l{\o}y}}, \bibinfo {author} {\bibfnamefont {A.}~\bibnamefont
  {Hansen}}, \ and\ \bibinfo {author} {\bibfnamefont {G.~G.}\ \bibnamefont
  {Batrouni}},\ }\href {\doibase https://doi.org/10.1023/A:1006510106194}
  {\bibfield  {journal} {\bibinfo  {journal} {Transport in porous media}\
  }\textbf {\bibinfo {volume} {32}},\ \bibinfo {pages} {163} (\bibinfo {year}
  {1998})}\BibitemShut {NoStop}%
\bibitem [{\citenamefont {Knudsen}\ \emph {et~al.}(2002)\citenamefont
  {Knudsen}, \citenamefont {Aker},\ and\ \citenamefont {Hansen}}]{kah02}%
  \BibitemOpen
  \bibfield  {author} {\bibinfo {author} {\bibfnamefont {H.~A.}\ \bibnamefont
  {Knudsen}}, \bibinfo {author} {\bibfnamefont {E.}~\bibnamefont {Aker}}, \
  and\ \bibinfo {author} {\bibfnamefont {A.}~\bibnamefont {Hansen}},\ }\href
  {\doibase https://doi.org/10.1023/A:1015039503551} {\bibfield  {journal}
  {\bibinfo  {journal} {Transport in Porous Media}\ }\textbf {\bibinfo {volume}
  {47}},\ \bibinfo {pages} {99} (\bibinfo {year} {2002})}\BibitemShut {NoStop}%
\bibitem [{\citenamefont {T{\o}r{\aa}}\ \emph {et~al.}(2012)\citenamefont
  {T{\o}r{\aa}}, \citenamefont {{\O}ren},\ and\ \citenamefont
  {Hansen}}]{toh12}%
  \BibitemOpen
  \bibfield  {author} {\bibinfo {author} {\bibfnamefont {G.}~\bibnamefont
  {T{\o}r{\aa}}}, \bibinfo {author} {\bibfnamefont {P.-E.}\ \bibnamefont
  {{\O}ren}}, \ and\ \bibinfo {author} {\bibfnamefont {A.}~\bibnamefont
  {Hansen}},\ }\href {\doibase https://doi.org/10.1007/s11242-011-9895-6}
  {\bibfield  {journal} {\bibinfo  {journal} {Transport in Porous Media}\
  }\textbf {\bibinfo {volume} {92}},\ \bibinfo {pages} {145} (\bibinfo {year}
  {2012})}\BibitemShut {NoStop}%
\bibitem [{\citenamefont {Gjennestad}\ \emph {et~al.}(2018)\citenamefont
  {Gjennestad}, \citenamefont {Vassvik}, \citenamefont {Kjelstrup},\ and\
  \citenamefont {Hansen}}]{gvkh18}%
  \BibitemOpen
  \bibfield  {author} {\bibinfo {author} {\bibfnamefont {M.~A.}\ \bibnamefont
  {Gjennestad}}, \bibinfo {author} {\bibfnamefont {M.}~\bibnamefont {Vassvik}},
  \bibinfo {author} {\bibfnamefont {S.}~\bibnamefont {Kjelstrup}}, \ and\
  \bibinfo {author} {\bibfnamefont {A.}~\bibnamefont {Hansen}},\ }\href
  {\doibase https://doi.org/10.3389/fphy.2018.0005} {\bibfield  {journal}
  {\bibinfo  {journal} {Frontiers in Physics}\ }\textbf {\bibinfo {volume}
  {6}},\ \bibinfo {pages} {56} (\bibinfo {year} {2018})}\BibitemShut {NoStop}%
\bibitem [{\citenamefont {Sinha}\ \emph {et~al.}(2011)\citenamefont {Sinha},
  \citenamefont {Gr{\o}va}, \citenamefont {{\O}deg{\aa}rden}, \citenamefont
  {Skjetne},\ and\ \citenamefont {Hansen}}]{sinhaGrova2011local}%
  \BibitemOpen
  \bibfield  {author} {\bibinfo {author} {\bibfnamefont {S.}~\bibnamefont
  {Sinha}}, \bibinfo {author} {\bibfnamefont {M.}~\bibnamefont {Gr{\o}va}},
  \bibinfo {author} {\bibfnamefont {T.~B.}\ \bibnamefont {{\O}deg{\aa}rden}},
  \bibinfo {author} {\bibfnamefont {E.}~\bibnamefont {Skjetne}}, \ and\
  \bibinfo {author} {\bibfnamefont {A.}~\bibnamefont {Hansen}},\ }\href
  {\doibase https://doi.org/10.1103/PhysRevE.84.037303} {\bibfield  {journal}
  {\bibinfo  {journal} {Physical Review E}\ }\textbf {\bibinfo {volume} {84}},\
  \bibinfo {pages} {037303} (\bibinfo {year} {2011})}\BibitemShut {NoStop}%
\bibitem [{\citenamefont {Flovik}\ \emph {et~al.}(2015)\citenamefont {Flovik},
  \citenamefont {Sinha},\ and\ \citenamefont
  {Hansen}}]{flovikSinha2015dynamic}%
  \BibitemOpen
  \bibfield  {author} {\bibinfo {author} {\bibfnamefont {V.}~\bibnamefont
  {Flovik}}, \bibinfo {author} {\bibfnamefont {S.}~\bibnamefont {Sinha}}, \
  and\ \bibinfo {author} {\bibfnamefont {A.}~\bibnamefont {Hansen}},\ }\href
  {\doibase https://doi.org/10.3389/fphy.2015.00086} {\bibfield  {journal}
  {\bibinfo  {journal} {Frontiers in Physics}\ }\textbf {\bibinfo {volume}
  {3}},\ \bibinfo {pages} {86} (\bibinfo {year} {2015})}\BibitemShut {NoStop}%
\bibitem [{\citenamefont {Blunt}(2017)}]{blunt2017multiphase}%
  \BibitemOpen
  \bibfield  {author} {\bibinfo {author} {\bibfnamefont {M.~J.}\ \bibnamefont
  {Blunt}},\ }\href {\doibase https://doi.org/10.1017/9781316145098} {\emph
  {\bibinfo {title} {Multiphase flow in permeable media: A pore-scale
  perspective}}}\ (\bibinfo  {publisher} {Cambridge University Press},\
  \bibinfo {year} {2017})\BibitemShut {NoStop}%
\bibitem [{\citenamefont {Straley}(1977)}]{s77}%
  \BibitemOpen
  \bibfield  {author} {\bibinfo {author} {\bibfnamefont {J.~P.}\ \bibnamefont
  {Straley}},\ }\href {\doibase 10.1103/PhysRevB.15.5733} {\bibfield  {journal}
  {\bibinfo  {journal} {Physical Review B}\ }\textbf {\bibinfo {volume} {15}},\
  \bibinfo {pages} {5733} (\bibinfo {year} {1977})}\BibitemShut {NoStop}%
\bibitem [{\citenamefont {Xun}\ \emph {et~al.}(2021)\citenamefont {Xun},
  \citenamefont {Hao},\ and\ \citenamefont {Ziff}}]{xhz21}%
  \BibitemOpen
  \bibfield  {author} {\bibinfo {author} {\bibfnamefont {Z.}~\bibnamefont
  {Xun}}, \bibinfo {author} {\bibfnamefont {D.}~\bibnamefont {Hao}}, \ and\
  \bibinfo {author} {\bibfnamefont {R.~M.}\ \bibnamefont {Ziff}},\ }\href
  {\doibase 10.1103/PhysRevE.103.022126} {\bibfield  {journal} {\bibinfo
  {journal} {Phys. Rev. E}\ }\textbf {\bibinfo {volume} {103}},\ \bibinfo
  {pages} {022126} (\bibinfo {year} {2021})}\BibitemShut {NoStop}%
\bibitem [{\citenamefont {Cen}\ \emph {et~al.}(2012)\citenamefont {Cen},
  \citenamefont {Liu},\ and\ \citenamefont {Mao}}]{wlm12}%
  \BibitemOpen
  \bibfield  {author} {\bibinfo {author} {\bibfnamefont {W.}~\bibnamefont
  {Cen}}, \bibinfo {author} {\bibfnamefont {D.}~\bibnamefont {Liu}}, \ and\
  \bibinfo {author} {\bibfnamefont {B.}~\bibnamefont {Mao}},\ }\href {\doibase
  10.1016/j.physa.2011.01.003} {\bibfield  {journal} {\bibinfo  {journal}
  {Physica A: Statistical Mechanics and its Applications}\ }\textbf {\bibinfo
  {volume} {391}},\ \bibinfo {pages} {925} (\bibinfo {year}
  {2012})}\BibitemShut {NoStop}%
\bibitem [{\citenamefont {Den~Nijs}(1979)}]{n79}%
  \BibitemOpen
  \bibfield  {author} {\bibinfo {author} {\bibfnamefont {M.}~\bibnamefont
  {Den~Nijs}},\ }\href {\doibase 10.1088/0305-4470/12/10/030} {\bibfield
  {journal} {\bibinfo  {journal} {Journal of Physics A: Mathematical and
  General}\ }\textbf {\bibinfo {volume} {12}},\ \bibinfo {pages} {1857}
  (\bibinfo {year} {1979})}\BibitemShut {NoStop}%
\bibitem [{\citenamefont {Erpelding}\ \emph {et~al.}(2013)\citenamefont
  {Erpelding}, \citenamefont {Sinha}, \citenamefont {Tallakstad}, \citenamefont
  {Hansen}, \citenamefont {Flekk{\o}y},\ and\ \citenamefont
  {M{\aa}l{\o}y}}]{esthfm13}%
  \BibitemOpen
  \bibfield  {author} {\bibinfo {author} {\bibfnamefont {M.}~\bibnamefont
  {Erpelding}}, \bibinfo {author} {\bibfnamefont {S.}~\bibnamefont {Sinha}},
  \bibinfo {author} {\bibfnamefont {K.~T.}\ \bibnamefont {Tallakstad}},
  \bibinfo {author} {\bibfnamefont {A.}~\bibnamefont {Hansen}}, \bibinfo
  {author} {\bibfnamefont {E.~G.}\ \bibnamefont {Flekk{\o}y}}, \ and\ \bibinfo
  {author} {\bibfnamefont {K.~J.}\ \bibnamefont {M{\aa}l{\o}y}},\ }\href
  {\doibase 10.1103/PhysRevE.88.053004} {\bibfield  {journal} {\bibinfo
  {journal} {Physical Review E}\ }\textbf {\bibinfo {volume} {88}},\ \bibinfo
  {pages} {053004} (\bibinfo {year} {2013})}\BibitemShut {NoStop}%
\end{thebibliography}%
\end{document}